  \documentstyle[prc,aps]{revtex}
\begin{document}
 \draft
 \title{\bf Competing electric and magnetic excitations \\ 
 in backward electron scattering \\ from heavy deformed nuclei}
\author{R. Nojarov,\cite{adno}  Amand Faessler\cite{adfa} 
  and M.~Dingfelder\cite{addi}}
\address{
Institut f\"ur Theoretische Physik, Universit\"at T\"ubingen, \\
Auf der Morgenstelle 14, D-72076 T\"ubingen, Germany}
\date{\today}
\maketitle
 \begin{abstract}
Important $E2$ contributions to the $(e,e^{\prime})$ cross
sections of low-lying orbital $M1$ excitations are found in heavy
deformed nuclei, arising from the small energy separation between the
two excitations with $I^{\pi}K = 2^+1$ and 1$^+1$, respectively. 
They are studied microscopically in QRPA using DWBA.  The
accompanying $E2$ response is negligible at small momentum transfer
$q$ but contributes substantially to the cross sections measured at
$\theta = 165 ^{\circ}$ for $0.6 < q_{\rm eff} < 0.9$ fm$^{-1}$ ($40
\le E_i \le 70$ MeV) and leads to a very good agreement with
experiment. The electric response is of longitudinal $C2$ type for 
$\theta \le 175 ^{\circ}$ but becomes almost purely transverse $E2$
for larger backward angles. The transverse $E2$ response remains 
comparable with the $M1$ response for $q_{\rm eff} > 1.2$ fm$^{-1}$ 
($E_i > 100$ MeV) and even dominant for $E_i > 200$ MeV. This happens
even at large backward angles $\theta > 175 ^{\circ}$, where the $M1$
dominance is limited to the lower $q$ region. 
 \end{abstract}

 \twocolumn 
\section{Motivation}

Inelastic electron scattering at backward angle $\theta =
180^{\circ}$ was used by Barber, Peterson {\sl et al.}
\cite{pet,bar63} more than 30 years ago to study nuclear magnetic
dipole ($M1$) excitations at the linear accelerator in Stanford.  The
predominance of $M1$ excitations in backward scattering is expected
from qualitative considerations \cite{bar63} of the $(e,e^{\prime})$
cross section in the plane wave Born approximation (PWBA). The cross
section can be decomposed in this case into longitudinal (Coulomb)
and transverse (electric and magnetic) terms, multiplied by
corresponding kinematical factors. The electric field of the incoming
electron can have both longitudinal and transverse components, while
its magnetic field is purely transverse. If the electron rest mass
and the nuclear excitation energy can be neglected in comparison with
the incident electron energy, which is often the case with low-lying
excitations, the longitudinal kinematical factor vanishes for $\theta
= 180^{\circ}$ and only transverse multipoles are excited by the
inelastic scattering \cite{bar62}.

  The $(e,e^{\prime})$ cross section can be related further to the
transition probability for photoabsorption through approximate
considerations within the "virtual photon" method. One arrives in
this way to the rough qualitative estimate \cite{bar63} for dominant
magnetic over electric transverse excitations (both of the same
multipolarity). This property is due to the momentum transfer during
electron scattering. It stands in contrast with the corresponding
relationship for radiative excitation, where no momentum is
transferred and the $EL$ excitation is one order of magnitude
stronger than the $ML$ excitation ($L$ stands for multipolarity).

   A more definite qualitative conclusion about the magnetic
dominance at backward angle can be drawn by assuming a purely
spin-flip transition \cite{fagg}. The transverse electric field of
the incoming electron is negligibly small in this particular case if
the excitation energy $E_x$ is small in comparison with the
transferred momentum $q$, i.e. $E_x \ll q \hbar c$. The same result
is obtained for a spin-flip transition in the limit of small momentum
transfer, $q \to E_x/\hbar c$ \cite{uber}. All these considerations
are made in the long-wave limit, $qR \ll 1$, where $R$ is the radius
of the target nucleus. The momentum transfer has throughout this
article the inverse dimension of length (fm$^{-1}$), i.e., $q/\hbar$
is denoted by $q$. 

   The peculiar selectivity of the backward  $(e,e^{\prime})$
scattering towards magnetic excitations was used extensively in the
last decades to study $M1$ transitions mainly in spherical nuclei
\cite{fagg,ramfagg}. The $M1$ operator can excite there particle-hole
states, which are spin-orbit partners. These excitations are
therefore often called spin-flip, though the orbital part of the $M1$
operator has also non-vanishing matrix elements \cite{noj95}. In
addition to these predominantly spin transitions, a qualitatively new
type of mainly orbital $M1$ excitations with $I^{\pi}K = 1^+1$ become
possible in deformed nuclei \cite{noj95}. They are closely related
to the so-called scissors mode, predicted by the two-rotor model
\cite{trm,iudice}. It describes isovector rotational oscillations of
neutrons against protons, which exist only in deformed nuclei.  Their
experimental study started a decade ago through backward inelastic
electron scattering at the linear accelerator in Darmstadt, see e.g.
the review articles \cite{achimrev}.

   We have calculated within the quasiparticle random-phase
approximation (QRPA) the  $(e,e^{\prime})$ form factors of low-lying
orbital $M1$ excitations in Titanium \cite{nofalip}, rare-earth
\cite{fanos}, and actinide \cite{noj93} nuclei. The present study was
motivated firstly by noticing at higher momentum transfer some
discrepancies between our theoretical $M1$ form factors and
experimental data. These deviations are not present in lighter nuclei
but seemed to occur systematically in heavy nuclei.  The first
diffraction maximum of the experimental $(e,e^{\prime})$ cross
section is usually well described theoretically when the experimental
$B(M1)$ value is reproduced in QRPA. In contrast, the $q$-dependence of
the theoretical cross section becomes very sensitive to the details
of the microscopic wave function at higher momentum transfer $q$. 

Our theoretical $M1$ transition density of the 1$^+$ excitation at
3.78 MeV in $^{48}$Ti \cite{nofalip} describes well the corresponding
experimental form factor \cite{guhr} not only at low but also up to
the highest measured momentum transfer, $q_{\rm eff} \le 2$
fm$^{-1}$.  We were even able to predict \cite{nofalip} the form
factor of the strongest $M1$ excitation (at 7.2 MeV) in this nucleus
in the same $q$-range before the corresponding experimental data
became available \cite{achim90}. The comparison with experiment can
be seen in Fig.~3 of Ref. \cite{noj95}.

  On the other hand, our theoretical $M1$ form factors for rare-earth
nuclei \cite{fanos}, though lying within the experimental error bars,
seem to underestimate systematically the experimental values even at
lower momentum transfer (for $q > 0.6$ fm$^{-1}$). In contrast to
$^{48}$Ti, high-$q$ experimental data are not available for single
$M1$ excitations in heavy deformed nuclei.  We were inclined to
attribute initially the above discrepancies to the adopted
parametrization of the Woods-Saxon potential. The $K^{\pi}=1^+$
excitations in deformed nuclei are weakly collective
\cite{nofa88,nfsciss}, because the main collectivity is concentrated
in the spurious isoscalar rotational state. Thus, the microscopic
wave functions of the low-lying orbital $M1$ excitations are very
sensitive to the details of the single-particle level scheme around
the Fermi surface. In order to improve the agreement with the
experimental form factors, we decided to shift in rare-earth nuclei
the energy of several single-particle levels close to the nuclear
surface \cite{fanos}, a procedure used formerly in microscopic
calculations for spherical nuclei. The modified level scheme did not
produce, however, a substantial improvement.

  Thus, we turn our attention to possible contributions from other
multipolarities, which should have the property of appearing only in
heavy nuclei. We are going to present in the next section preliminary
qualitative estimates for the expectation of $E2$ contributions to
the $(e,e^{\prime})$ cross sections at backward angle. The
microscopic calculation of transition densities within our QRPA
formalism is described in Sect. III. The theoretical $(e,e^{\prime})$
cross sections, obtained from these transition densities in the
distorted wave Born approximation (DWBA) for $\theta = 165 ^{\circ}$,
are compared to experiment in Sect.  IV. A qualitative analysis of
our DWBA results for larger backward angles is presented in Sect. V
using more transparent PWBA estimates. The conclusions are
summarized in Sect. VI. The most important ingredients of our
formalism can be found in Appendices A-D. 

\section{Preliminary estimates}

Our search for possible contributions from higher multipolarities was
stimulated by the experimental study \cite{bohle85} of the $E2$
transition accompanying the strongest $M1$ excitation, $E_x$ = 3.075
MeV, in $^{156}$Gd. The $E2$ transition excites the first member
(with $I^{\pi}K = 2^+1$) of the rotational band built on the top of
the vibrational state with $I^{\pi}K = 1^+1$.  This is the only
accompanying $E2$ transition studied until now in heavy deformed
nuclei. The energy separation between the two excitations ($M1$ and
$E2$) corresponds to the rotational energy
\begin{equation}
\Delta E_{12} =  { [ I(I+1) - K(K+1) ] \hbar ^2 \over 2 {\cal J} } 
=  {2 \hbar ^2 \over {\cal J} }, 
\label{roten} \end{equation}
where ${\cal J}$ is the moment of inertia of the rotational band
developed on the intrinsic 1$^+$ state. Since its value is not known,
one could use as a rough estimate the ground-state moment of inertia,
determined from the energy $E(2^+_1) = 3 \hbar ^2 / {\cal J}_{g.s.}$
of the first rotational 2$^+$ state in $^{156}$Gd \cite{raman}. The
so obtained energy separation, $\Delta E_{12} = 59.3$ keV, should be
considered as an upper limit, because the rotational bands of excited
states have usually larger moments of inertia than the g.s. band
\cite{nona}. The 2$^+$1 state found in $^{156}$Gd \cite{bohle85} lies
indeed only 21 keV above its 1$^+$1 band head with $E_x = 3.075$ MeV.
This energy separation is comparable with the energy resolution
$\Delta E_{exp.}$ of the considered $(e,e^{\prime})$ experiments, 20
keV $ < \Delta E_{exp.} < $ 40 keV, depending on the incident
electron energy \cite{bohle86}.
 
   The largest backward angle was $\theta = 165^{\circ}$ in these
experiments.  It corresponds to a ratio between the transverse and
longitudinal kinematic factors \cite{fagg}
\begin{equation}
 { V_t \over V_{\ell} } = {1 \over 2} + {\rm tg}^2 \left ( 
 {\theta \over 2} \right ) = 58.  \label{vtrat} 
\end{equation} 
Although the longitudinal Coulomb excitation is strongly damped by
this large factor, it could still provide a $C2$ contribution to the
cross section comparable with the $M1$ contribution. This is due to the
fact that the longitudinal $E2$ transition matrix element, before being
reduced by the above factor, is usually much larger than the $M1$
matrix element.  A new 180$^ {\circ}$ electron scattering facility is
already in operation in Darmstadt \cite{darm180}. Though the aperture
of such facilities leads to a somewhat smaller effective mean angle
of 178.5$^{\circ}$ \cite{petra}, the longitudinal contributions are
already negligible at this angle, where the ratio (\ref{vtrat})
amounts to 5848. On the other hand, the worsened energy resolution,
typically $\Delta E_{exp.} = 100$ keV at a fully backward angle
\cite{petra}, makes quite plausible the appearance of transverse $E2$
contributions to the measured $M1$ cross section, if both are of
comparable magnitude.
 
The ground state of even-even nuclei corresponds to $I^{\pi}K =
0^+0$, so that due to the selection rules each one of the above
considered 1$^+$1 and 2$^+$1 states can be excited only by $M1$ and
$E2$ transitions, respectively, but not by both of them.  However,
the multipolarity of the  excited states studied is usually not known
in experiment. Thus, most of the former qualitative estimates for the
relative $CL, \; EL$, and $ML$ contributions to the PWBA cross
section consider transitions with the same multipolarity $L$ in order
to find out what kinematics would enhance selectively the $ML$
excitation.

In contrast, we are interested in the present study in the relative
$E2$ and $M1$ contributions. They excite two different states, but
contribute nevertheless to the experimentally measured cross section
only because in heavy deformed nuclei the energy separation of the
two states is comparable with the experimental energy resolution.
This situation does not occur in lighter even-even nuclei whose small
moment of inertia produces a large energy separation between the two
states and the two cross sections are easily resolved in experiment.
We denote by $E2$ both longitudinal $C2$ and transverse $E2$
components, unless one of them is specified explicitly.

   Our study was motivated also by one of the first calculations of
$(e,e^{\prime})$ cross sections in heavy deformed nuclei
\cite{tassie}.  The PWBA cross sections for members of the
ground-state and $\beta$-vibrational bands in $^{152}$Sm are
obtained in the collective Tassie model assuming an incompressible
and irrotational fluid.  Although a purely longitudinal (Coulomb)
excitation is described by this simple model, the qualitative
$q$-behaviour of the calculated cross sections supports our
expectations of improving the $M1$ cross sections by taking $E2$
contributions into account: i) the diffraction maxima of higher
multipolarities have smaller amplitudes but are broader, ii) the
$I^{\pi}K = 2^+0$ state of the $\beta$-vibrational band  has even a
larger cross section than the band head (0$^+$0), i.e. the magnitude
ordering is inversed for the intrinsic excitation with respect to the
g.s. band. These qualitative features were confirmed experimentally
some 20 years later \cite{hers} through $(e,e^{\prime})$ on
$^{154}$Gd.

  The above arguments about possible appreciable $E2$
electroexcitations even at backward angles stimulated us
to examine more carefully the qualitative estimates, on which the
commonly expected $M1$ dominance in backward electron scattering is 
based. The relative contribution of the transverse magnetic and
electric terms to the $(e,e^{\prime})$ cross section is given by the
approximate PWBA relationship \cite{bar63}: 
\begin{equation}
{ d\sigma (ML) \over d\sigma (EL)} \approx { 1 \over 10} \left ( 
{q \hbar c \over E_x} \right ) ^2, \label{barb} 
\end{equation}
where we have used the correct ratio from Eq. (30) of Ref.
\cite{bar62}. The factor 10 is a rough estimate for the ratio between
the two cross sections at the photon point where they are
proportional to the respective reduced probabilities for radiative
transitions.  Thus, one should expect predominantly magnetic
excitations in backward electron scattering, because in most
experiments $E_x \ll q\hbar c$. But the qualitative relationship
(\ref{barb}) was derived within the "virtual photon" method valid for
$q \to E_x/\hbar c$. It was stated explicitly \cite{bar63} that the
relationship (\ref{barb}) is not enough to upset in general the
dominant magnetic character of backward electron scattering.  More
detailed considerations will be discussed in Sect.\  V below. 

   Without recurring to virtual photons, Willey arrives to a more
general qualitative relationship between the two cross sections
\cite{willey}: 
\begin{equation}
{ d\sigma (ML) \over d\sigma (EL)} \approx { 1 \over \left ( 
qR \right ) ^2 }. \label{willey} 
\end{equation}
The magnetic transition is obviously dominant, because
(\ref{willey}) is derived in the long-wave limit, $qR \ll 1$. 
However, the long-wave approximation is not fulfilled in the 
above discussed experiments with heavy nuclei
\cite{achimrev,bohle86}, where $q > 0.2$ fm$^{-1}, \quad R \approx
6.5$ fm, and $qR > 1.3$. With this numerical value one could
conclude from (\ref{willey}) that the transverse electric should be
comparable with the transverse magnetic electroexcitation, and even
dominant for $q > 0.5$ fm$^{-1}$, i.e.  in the region where we
expect important $E2$ contributions. Of course, (\ref{willey}) is no
more valid for such large values of $qR$.

   The above simple considerations are not able to provide a 
reliable prediction for the relative interplay of electric and
magnetic electroexcitations in backward scattering. They ignore
the dynamics contained in the wave functions of the studied nuclear
excitations. More definite conclusions can be drawn only from
realistic microscopic calculations which are able to provide
satisfactory quantitative description of the experimental data. We
are using the DWBA because the large charge of heavy nuclei produces
a strong Coulomb potential distorting the electromagnetic field of
the incoming electron.  After our first encouraging results on
$^{156}$Gd \cite{nfd94,dnf95}, we are going to study here the problem
of electric contributions to backward scattering more systematically
and to compare our theoretical results with all the available
experimental cross sections for single $M1$ excitations in rare-earth
nuclei.
 
\section{Transition densities in QRPA}

   We solve the QRPA equations of motion for  intrinsic excitations
with $K^{\pi} =1^+$ using a model hamiltonian  \cite{nfd95},  
\begin{eqnarray}
 \hat{H} = \hat{H}_0 + \hat{H}_{\rm FF} + \hat{H}_{\rm SS},
\nonumber \\
\hat{F}(m,t) = [ \hat{H}_0, \hat{J} (m,t) ], \quad t = n, \, p, 
\quad r = { k_1 \over k_0}, \label{ham} 
\end{eqnarray} 
where the isospin index $t=n,p$ denotes neutrons or protons.  The
quasiparticle (q.p.) mean field $\hat{H}_0$ \cite{nofa88,fano90} is
given by an axially-symmetric Woods-Saxon potential \cite{dam69} plus
BCS pairing. Apart from the deformation, all the remaining
Woods-Saxon parameters are given for each nucleus by the
isospin dependent parametrization of Ref. \cite{tanaka}. It is based
on a single set of isospin independent parameters for all nuclei
without respect to their mass and charge numbers. The hexadecapole
deformation is taken from experiment \cite{beta4}, while the
quadrupole one is determined for each nucleus by fitting the
microscopically calculated ground state quadrupole moment to the
corresponding experimental value \cite{ram87}. As in most of our
previous works, we use an oscillator constant $\hbar \omega _0 = 50
\, A^{-1/3}$ MeV and include in the basis (\ref{basis}) all the
oscillator eigenfunctions with principal quantum numbers 
(\ref{qnum}) ${\cal N} \le 11$. The pairing gaps are chosen to
reproduce roughly the even-odd mass differences
\cite{wapstra}.

  The next two terms in the model hamiltonian (\ref{ham}) are
separable residual interactions of quadrupole and spin type,
respectively. The former interaction is constructed with operators of
quadrupole type $\hat{F}$ (\ref{ham}) derived from the deformed part
of the q.p. mean field \cite{nofa88,nfd95}. The total angular
momentum $\hat{J} (m,t)$ (\ref{mop}) is symmetrized over the
signature $m=\pm 1$, see Appendix B. The isoscalar coupling constant
$k_0$ of $\hat{H}_{\rm FF}$ is determined microscopically
\cite{baznat} in order to restore the rotational invariance of the
model hamiltonian, violated by the deformation. The value $r = -2$ is
adopted for the ratio (\ref{ham}) between the isovector and the
isoscalar coupling constants. This choice, discussed in \cite{nfd95},
gives rise in QRPA to a high-lying $E2$ strength, whose energy
distribution reproduces qualitatively the isovector giant quadrupole
resonance observed in $(e,e^{\prime})$ experiments.  We adopt the
above rough round value, because of the uncertainties of the scarce
experimental data on heavy deformed nuclei. Our attention is focused
here on the low-lying orbital $M1$ excitations, which are not very
sensitive towards the strength of the isovector quadrupole
interaction. They are influenced much stronger by the exclusion of
spuriousity through restoration of the rotational invariance. They
acquire afterwards a well pronounced character of isovector
rotational vibrations \cite{nofa88,nfsciss}.

   The last term in (\ref{ham}) is the spin-spin interaction 
$\hat{H}_{\rm SS}$ \cite{fano90}. Its coupling constants, 
\begin{equation}
c(+)A = 200 \; \hbox{MeV}, \quad c(-) = -0.5 c(+), \label{sc} 
\end{equation}
are derived from nuclear matter calculations using the Reid soft core
interaction in the way explained in \cite{noj93}. The coupling
strengths (\ref{sc}) provide a good description \cite{sarri} of the
experimental spin $M1$ strength in rare-earth nuclei, deduced from
inelastic proton scattering.

  We present here results for all the rare-earth nuclei with
experimentally measured  $(e,e^{\prime})$ cross sections of single
1$^+$ excitations. They are characterized by a relatively strong $M1$
transition and an excitation energy of about 3 MeV.  The energies
$E_x$ of the $M1$ excitations, the $B(M1)$ and $B(E2)$ values, and
the orbit-to-spin ratio $R_{\rm o.s.}$ of the $M1$ matrix elements
are compared with experiment in Table \ref{tabrpa}.  Expressions for
$R_{\rm o.s.}$, $M1$ and $E2$ transition probabilities in terms of
the RPA amplitudes are given in \cite{nofa88}. The $B(M1)$ values for
transitions to $I^{\pi}K =1^+1$ states are calculated with bare
orbital and effective spin gyromagnetic factors, $g_s = 0.7 g_s^{\rm
free}$. The $B(E2)$ values for transitions to the corresponding
$I^{\pi}K =2^+1$ states, discussed already in Sect. II, are obtained
without introducing effective charges, i.~e. they are of purely
proton nature. The same orbital and spin gyromagnetic factors and
effective charges are used, of course, in the calculation of
transition densities, presented below.

It is seen from Table \ref{tabrpa} that the theoretical results
are in a good agreement with the available experimental data from
$(e,e^{\prime})$, $(p,p^{\prime})$, and $(\gamma ,\gamma^{\prime})$
scattering. One should note especially the agreement with the
experimental $B(E2)$ value for $^{156}$Gd, extracted from
$(e,e^{\prime})$ \cite{bohle85}. It is closely connected with the
main topic of the present study: the $E2$ response accompanying the
$M1$ electroexcitations in backward scattering. Moreover, as
discussed in Sect.\ II, this is the only accompanying $E2$ transition
identified experimentally as yet.

 The experimental situation is more complicated in $^{158}$Gd, where
the $(e,e^{\prime})$ peak \cite{boh84} at 3.2 MeV with $B(M1) =
1.4(3) \; \mu^2_N$ is resolved in nuclear resonance fluorescence
\cite{pitz89} in two $M1$ transitions: the first one is at 3.19 MeV
with $B(M1) = 0.66(8) \; \mu^2_N$, and the second one is listed in
Table \ref{tabrpa}. As discussed in Sect. II, each $M1$ transition is
accompanied by a corresponding $E2$ excitation. The latter was
studied experimentally only in $^{156}$Gd as yet, where it has an
excitation energy $E_x (2^+1) = 3.096$ MeV \cite{bohle85}.

The large ratio $R_{\rm o.s.}$ \cite{nofa88} between the orbital and
spin $M1$ matrix elements of each excitation, listed in the last
column of Table \ref{tabrpa}, indicates that these are predominantly
orbital transitions. Most of them represent the strongest single $M1$
transition in the corresponding nucleus. They were analysed by
comparison with the microscopic realization \cite{fanota} of the
scissors mode \cite{trm}, as well as with synthetic rotational phonons
\cite{fanos}. This analysis allows us to interpret them as a weakly
collective scissors mode \cite{nofa88,nfsciss,fanos}. 

   Electroexcitations are described by the following one-body
charge and current density operators \cite{willey,uber}: 
\begin{eqnarray}
{\hat \rho} ({\bf r}) = e \sum _{j=1} ^A \varepsilon _j \delta 
({\bf r} - {\bf r}_j), \label{rhop} \\
{\hat {\bf J}} ({\bf r}) = {\hat {\bf j}} ^{\rm C} ({\bf r}) + 
{\hat {\bf j}} ^{\rm S} ({\bf r}),  \label{jtop} \\
{\hat {\bf j}} ^C ({\bf r}) = { e \over Mc} \sum _{j=1} ^A 
\varepsilon _j {\hat {\bf p}}_j  \delta ({\bf r} - {\bf r}_j), 
 \quad {\hat {\bf p}}_j = - i \hbar \hat{\bbox{\nabla}} _j, 
 \label{convop}  
 \end{eqnarray}
\begin{eqnarray}
 {\hat {\bf j}} ^{\rm S} ({\bf r}) = \bbox{\nabla} \times 
 \hat{\bbox{\mu}} ({\bf r}), \nonumber \\
 \hat{\bbox{\mu}} ({\bf r}) = {e \over 2Mc} \sum _{j=1} ^A 
 g^s_j {\hat {\bf S}}_j \; \delta ({\bf r} - {\bf r}_j),
 \label{spinop} \end{eqnarray}
where $e, M, \varepsilon$ and $g^s$ are the unit charge, the nucleon
mass, effective charges and spin gyromagnetic ratios, respectively.
${\hat \rho} ({\bf r})$ and ${\hat {\bf J}} ({\bf r})$ are scalar
charge density and vector current density operators, respectively.
The latter is a sum of the convection current ${\hat {\bf j}} ^{\rm
C} ({\bf r})$ and the magnetization (or spin) current ${\hat {\bf j}}
^{\rm S} ({\bf r})$. The summation runs over the coordinates of all
the nucleons.

The momentum operator ${\bf p}$ in (\ref{convop}) has to
be hermitian symmetrized in order to satisfy the continuity equation
\cite{willey}: ${\bf p} ^{\rm sym.} = \textstyle{1 \over 2}
({\bf p} + {\bf p} ^{\dag}).$ However, we do not need this
symmetrization in (\ref{convop}) because we work always with
signature symmetrized operators, defined below (\ref{mop}). They
possess even higher symmetries, listed in Table \ref{tabphas},
Appendix B, including the hermitian property required for the
convection current (\ref{convop}).

   In consistence with our QRPA calculations, we do not use effective
charges in (\ref{rhop},\ref{convop}), but the spin operator is
renormalized by a factor of 0.7, included in the spin gyromagnetic
ratios:
\begin{eqnarray}
 \varepsilon ^n =0, \; \varepsilon ^p =1, \quad g^s_j = 0.7 g^s_j 
 ({\rm free}), \nonumber \\
g^s_n ({\rm free}) = -3.8263, \quad  g^s_p ({\rm free}) = 5.5858. 
\label{grat} \end{eqnarray}
Thus, only protons contribute to the calculated transition densities
of charge and convection currents.

  The charge and current density operators (\ref{rhop},\ref{jtop})
can be expanded in spherical harmonics and vector spherical
functions, respectively \cite{varsh}:
\begin{eqnarray}
{\hat \rho} ({\bf r}) = \sum _{LM} {\hat \rho}_{LM} (r) Y ^*
_{LM} (\Omega ), \nonumber \\
{\hat {\bf J}} ({\bf r}) = \sum _{LL^{\prime}M} {\bf Y}^* 
_{LL^{\prime}M} (\Omega ) {\hat {\cal J}} _{LL^{\prime}M} (r),  
\nonumber \\
{\bf Y}_{LL^{\prime}M} (\Omega ) = \sum _{\mu \mu ^{\prime}} 
(L^{\prime} \mu, 1 \mu ^{\prime} \vert LM) Y_{L^{\prime} \mu} 
(\Omega ) {\bf e}_{\mu ^{\prime}}. \label{multex} 
\end{eqnarray}
The resulting multipole density operators have the form
\begin{eqnarray}
{\hat \rho}_{LM} (r) = \int Y_{LM} (\Omega ) {\hat \rho} ({\bf r}) \;
d \Omega , \nonumber \\
{\hat {\cal J}} _{LL^{\prime}M} (r) = \int {\bf Y} _{LL^{\prime}M} 
(\Omega ) \bullet  {\hat {\bf J}} ({\bf r}) \; d \Omega,
\nonumber \\
L^{\prime} = L, L \pm 1. \label{mdop}
\end{eqnarray}

  We are interested only in $M1$ and $E2$ electroexcitations. The former
are generated by the transverse $M1$ current density operator ${\hat
{\cal J}} _{11^{\prime}M} (r)$. The latter receive longitudinal
(Coulomb) contributions from the charge density operator ${\hat
\rho}_{2M} (r)$ and transverse contributions from the $E2$ current
density operators ${\hat {\cal J}} _{2L^{\prime}M} (r), \;
L^{\prime} = 1,3$. Only operators with $M = \pm 1$ contribute to the
considered intrinsic excitations with $K^{\pi} =1^+$.

   We use $m$-symmetrized operators in our QRPA formalism
\cite{nofa88}. This is necessary to close the algebra of the   
commutation relations. Such operators possess also definite
symmetries with respect to complex, hermitian, and time conjugation
by simply changing their phase under these transformations, see Table
\ref{tabphas} in Appendix B. We are able to ensure in this way the
hermiticity of the convection current (\ref{convop}),  to take
further symmetries of the microscopic expressions analytically into
account, and to reach the minimal set of irreducible terms.  The
signature $m=\pm 1$ corresponds to the intrinsic $x$- and $y$-axes,
which are indistinguishable in axially symmetric nuclei. Thus, final
expression for physical observables are independent of $m$.

  There are two main types of operators in the quasiboson formalism:  
longitudinal and transverse, according on their phase $\sigma =
\pm 1$ (\ref{sigma}) with respect to combined hermitian and time
conjugation, considered in Appendix B.  Operators of the same type
commute with each other in the quasiboson approximation
(\ref{comut}).  Non-commuting operators are, e.g., the quadrupole
operator ${\hat Q}(m)$ and the total angular momentum ${\hat J} (m)$
\cite{nofa93}, being of longitudinal and transverse type,
respectively.  We define here additionally $m$-symmetrized multipole
charge $\hat{\rho}_2 (m,r)$ and current ${\hat {\cal J}}
_{LL^{\prime}} (m,r)$ density operators using (\ref{mdop}):
\begin{eqnarray}
\hat{J} (m) = \textstyle{1 \over 2} [ m \hat{J}_+ - \hat{J} 
^{\dag}_+ ], \quad \hat{J}_{\pm} = \hat{J}_x  \pm i \hat{J}_y, 
\nonumber \\
\hat{Q} (m) = \textstyle{1 \over 2} [ m \hat{Q}_{21} + 
\hat{Q}^{\dag} _{21} ], \nonumber \\
\hat{\rho}_2 (m,r) = \textstyle{1 \over 2} [ m \hat{\rho} _{21} (r) 
+ \hat{\rho} ^{\dag} _{21} (r) ], \nonumber \\
{\hat {\cal J}} _{LL^{\prime}} (m,r) = \textstyle{1 \over 2} [ m
{\hat {\cal J}} _{LL^{\prime}1} (r) + {\hat {\cal J}} ^{\dag}
_{LL^{\prime}1} (r) ]. 
\label{mop} \end{eqnarray} 
It is seen from their quasiboson representation (\ref{bosrep}) that
the charge density is of longitudinal (quadrupole) type with $\sigma
= +1$, while the current densities are of transverse (momentum) type
with $\sigma = -1$. The definitions (\ref{mop}) ensure that the
quasiparticle (q.p.) matrix elements of all transition operators have
the same symmetries (\ref{qpsym}).

   The $(e,e^{\prime})$ transition densities are the reduced matrix
elements (r.m.e.) of the multipole density operators (\ref{mdop})
between the initial and final nuclear states \cite{uber}. We describe
here in QRPA intrinsic excitations with $K^{\pi} = 1^+$ in axially
symmetric even-even nuclei, where $K$ is the projection of the total
angular momentum $I$, on the intrinsic symmetry axis of the nucleus
and $\pi$ is the parity of the wave function. The states are
characterized by $I^{\pi}K, \, \nu$, where  $\nu = 1,2,...$ labels
the intrinsic QRPA excitations $\Gamma ^{\dag} _{\nu} (m) \vert
\rangle$ (\ref{phon}), obtained by solving the QRPA equations of
motions.  

The transition $TL$ with multipolarity $L$ ($M1$ or $E2$) takes place
between the ground state with $I^{\pi}K=0^+0$ and an excited state
$\nu$ with  $I^{\pi}K=L^+1$.  The transition densities (\ref{rmel}),
corresponding to the multipole operators (\ref{mdop}), are obtained
from (\ref{intme}). We use the final expression (\ref{redsum}) in
terms of q.p. matrix elements (\ref{stepme}) of step-up operators,
because the microscopic sum (\ref{redsum}) contains the minimal set
of independent terms, not related with each other through any
symmetry. Thus, the $E2$ and $M1$ transition densities for the
$\nu$-th QRPA excitation have the form:
\begin{mathletters} \label{den} 
\begin{eqnarray}
\rho _2 ^{\nu} (r) = \langle 2^{\pi}1, \nu \parallel \hat{\rho} 
^{\dag}_2 (r) \parallel 0 \rangle = {1 \over 2} \sum _{0 < i < f} 
\nonumber \\
  \Bigl [ F^{\nu} _{+1} (fi) \rho _{21} (fi,r) +  F^{\nu} _{+1} 
(f\tilde{i}) \rho _{21} (f\tilde{i},r) \Bigr ], 
\label{rhoden} \end{eqnarray}
\begin{eqnarray}
{\cal J} _{LL^{\prime}}^{\nu} (r) = i \langle L^{\pi}1, \nu \parallel 
\hat{\cal J} ^{\dag} _{LL^{\prime}} (r) \parallel 0 \rangle = 
 {i \over 2} \sum _{0 < i < f} \nonumber \\
 \Bigl [ F^{\nu} _{-1} (fi) {\cal J} _{LL^{\prime}1} (fi,r)
+  F^{\nu} _{-1} (f\tilde{i}) {\cal J} _{LL^{\prime}1} 
(f\tilde{i},r) \Bigr ],  \nonumber \\
 {\cal J}^{\nu} _{LL^{\prime}} (r) =  {\cal J}^{\nu ,C}
_{LL^{\prime}}  (r) + {\cal J}^{\nu ,S} _{LL^{\prime}} (r), 
\label{jden} \end{eqnarray}
\end{mathletters}
\noindent
where ${\rho}_{21} (fi,r)$, ${\cal J} _{LL^{\prime}1} (fi,r)$, are
the q.p. matrix elements (\ref{rhome}), (\ref{jme}), (\ref{jcl}), and
(\ref{jsl}) of the transition density operators $\hat{\rho}_{21} (r)$
(longitudinal) and $\hat{\cal J} _{LL^{\prime}1} (r)$ (transverse),
given by (\ref{mdop}). The transverse transition density (\ref{jden})
is decomposed into a sum of convection and spin (or magnetization)
parts, according to (\ref{jtop})-(\ref{spinop}).  The information
about the considered $\nu$-th QRPA excitation is contained in the
factors $F^{\nu} _{\sigma} (fi)$, which are linear combinations
(\ref{fup}) of the QRPA amplitudes (\ref{phon}).
 
The QRPA transitions densities for $M1$ and $E2$ excitations of the 
$K^{\pi} = 1^+$ state in $^{156}$Gd from Table \ref{tabrpa} are
displayed in Fig.\ \ref{fig1}. This is the strongest low-lying
(orbital) $M1$ transition in $^{156}$Gd. The longitudinal charge
transition density $\rho _2 (r)$ has the largest amplitude (note the 
different ordinate scales), followed by the transverse $M1$
transition density ${\cal J}_{11} (r)$. Each transverse density
(continuous curve) is a sum of three components, arising from the
proton convection (long dashed curve) and spin (dotted curve for
protons and short dashed curve for neutrons) currents (\ref{jtop}).
In the case of ${\cal J}_{11} (r)$ and ${\cal J}_{21} (r)$ the proton
convection current is dominant, while the spin currents for protons
and neutrons are out-of-phase, as it should be expected for an
isovector vibration, and cancel each other to a great extent.

We do not display the transition densities of the other 1$^+$
excitations from Table\ \ref{tabrpa}, because they are very similar
to those in Fig.\ \ref{fig1}, apart from the cases of $^{158}$Gd
and $^{168}$Er. Since the latter two cases exhibit also similarities,
we show only the transition densities for $^{168}$Er in Fig.\
\ref{fig2}. This excitation has a more pronounced volume character
because, in contrast to Fig.\ \ref{fig1}, the transition densities
are no more peaked near the nuclear surface. Although the neutrons
have a large contribution to the RPA wave function of this
excitation, the neutron spin current is very small, as seen from the 
short-dashed curves in Fig.\ \ref{fig2}. 

\section{Interplay of $E2$ and $M1$ ($\lowercase{e}, \lowercase{e} 
^{\prime})$ cross sections}

We use the QRPA transition densities (\ref{den}) for a selected
intrinsic excitation 1$^+_{\nu}$ to calculate its $(e,e^{\prime})$
cross sections with the DWBA code of J.~Heisenberg \cite{heis}. The
$M1$ excitation is fully specified by the transverse transition
density ${\cal J} _{11} ^{\nu} (r)$, while the $E2$ excitation has
also a longitudinal (Coulomb) part, $\rho _2 ^{\nu} (r)$, besides the
two transverse components ${\cal J} _{2L^{\prime}} ^{\nu} (r), \>
L^{\prime} = 1,3$. The DWBA code uses only ${\cal J} _{23} ^{\nu}
(r)$ as input and takes the other transverse transition density
through the continuity equation into account. 

The DWBA $(e,e^{\prime})$ cross sections of the five 1$^+$
excitations from Table\ \ref{tabrpa} are plotted in Figs.\
\ref{fig3}-\ref{fig5} and compared to experiment. Each one of these
1$^+$ states corresponds to the strongest low-energy orbital $M1$
transition in the respective nucleus. These are all the cases in
which experimental $(e,e^{\prime})$ cross sections of single 1$^+$
excitations in rare-earth nuclei have been published until now. The
QRPA transition densities for $^{156}$Gd and $^{168}$Er, used in the
DWBA calculations, are shown in Figs.\ \ref{fig1} and \ref{fig2}.

The DWBA cross sections in Figs.\ \ref{fig3}-\ref{fig5} correspond to
the scattering angle $\theta = 165^{\circ}$, at which the experimental
data \cite{bohle84,hart89,boh84} (dots with error bars) were taken.
The wave of the incoming electron is distorted by the ground state
charge distribution of the target nucleus, described by a
two-parameter Fermi formula. We use for $^{154}$Sm and $^{156}$Gd the
experimental values \cite{vries87} of the two parameters, extracted
from electron scattering. Since no experimental values are available
for the other studied nuclei, the values for the latter nucleus are
used also for $^{154,158}$Gd, while the experimental values for
$^{166}$Er \cite{vries87} are employed for $^{168}$Er.

It is seen from Figs.\ \ref{fig3}-\ref{fig5} that the theoretical
$M1$ cross section of the $I^{\pi}K=1^+1$ excitation (dashed curve)
underestimates systematically the experimental data, especially at
higher momentum transfer $q_{\rm eff} > 0.6$ fm$^{-1}$, i.e., $q >
0.4$ fm$^{-1}$, corresponding to incident electron energy $E_i > 40$
MeV. This peculiarity was discussed already in the Motivation,
Sect.~I. At small momentum transfer (close to the photon point) the
cross section of the accompanying $E2$ transition to the
$I^{\pi}K=2^+1$ state (dotted curve) is two orders of magnitude
smaller than the $M1$ cross section. Thus, the $E2$ transition will
not introduce practically any correction to the experimental 
$B(M1)$ value, which is extracted by extrapolating the experimental
cross section to the photon point $q = E_x /\hbar c$. 

As discussed in Sect.\ II, the accompanying $E2$ response was
identified experimentally only in $^{156}$Gd as yet \cite{bohle85}.
Its experimental $B(E2)$ value is reproduced in our calculations
together with the corresponding $B(M1)$ value (Table \ref{tabrpa}).
This is an additional proof for the correct magnitudes at low $q$ of
our theoretical $E2$ and $M1$ DWBA cross sections for $^{156}$Gd,
displayed on the top plot of Fig.\ \ref{fig4}. The correct transition
probabilities provide, moreover, an additional test for our QRPA
transition densities $\rho _2 ^{\nu} (r)$ and ${\cal J} _{11} ^{\nu}
(r)$  from Fig.\ \ref{fig1} because of the simple relationships 
\cite{heis}:
\begin{eqnarray}
B(E2) = \Bigl [ \int _0 ^{\infty} \rho _2 ^{\nu} (r) r^4 dr 
\Bigr ]^2, \nonumber \\
B(M1) =  {1 \over 2} \Bigl [ \int _0 ^{\infty} {\cal J} _{11} ^{\nu} 
(r) r^3 dr \Bigr ]^2.  \label{intden}
\end{eqnarray}

The $E2$ and $M1$ cross sections become comparable in magnitude in
the region $0.7 < q_{\rm eff} < 0.8$ fm$^{-1}$ ($0.5 < q < 0.6$
fm$^{-1}, \; 50 \le E_i \le 60$ MeV), as seen from Figs.\
\ref{fig3}-\ref{fig5}.  The sum of the $E2$ and $M1$ theoretical
cross sections (continuous curve) agrees very well with the available
experimental data on $(e,e^{\prime})$ cross sections of single 1$^+$
excitations in rare-earth nuclei. The discrepancies in this
$q$-region, typical for the $M1$ cross section alone, have been
removed. It is seen from Figs.\ \ref{fig3}-\ref{fig5} that the
accompanying $E2$ transition provides an important contribution to
the measured cross sections for $0.6 < q_{\rm eff} < 0.9$ fm$^{-1} \>
(0.4 < q < 0.7$ fm$^{-1}, \; 40 \le E_i \le 70$ MeV). 

Let us consider further in more detail the 1$^+$ excitation in
$^{156}$Gd as an example. Its theoretical $M1$ cross section (dashed
line in the top plot of Fig.\ \ref{fig4}) is plotted once more as a
continuous curve in Fig.\ \ref{fig6} up to a higher momentum
transfer. It is obtained from the QRPA  transition density ${\cal J}
_{11} ^{\nu} (r)$ (continuous line in the corresponding plot of Fig.\
\ref{fig1}). If one neglects the spin current and uses only the
convection transition density ${\cal J}_{11} ^{\nu ,C} (r)$ (long
dashed line in Fig.\ \ref{fig1}), one obtains the DWBA cross section
displayed by a dashed curve in Fig.\ \ref{fig6}. 

The comparison between the continuous and dashed curves in Fig.\
\ref{fig6} shows that the $M1$ excitation has a predominantly
convection character, as it should be expected for an orbital
transition characterized by the large orbit-to-spin ratio $R_{\rm
o.s.}$ in Table \ref{tabrpa}. The total (convection plus spin) PWBA
cross section (dot-dashed curve in Fig.\ \ref{fig6}) is about one
order of magnitude smaller than the corresponding DWBA cross section
(continuous curve). Note that the PWBA cross section is plotted, as
usual, versus $q$, not $q_{\rm eff}$.

It is easier to draw qualitative conclusions in PWBA, where the
cross section can be decomposed into a sum of longitudinal and
transverse terms \cite{uber,heis}:
\begin{eqnarray}
\Bigl ( {d \sigma \over d \Omega} \Bigr )_{\rm PWBA} = \Bigl ( 
{Z e^2 \over E_i} \Bigr ) ^2 \Bigl \{ V_{\ell} \vert F^C_L (q) 
\vert ^2 \nonumber \\
+ V_t \bigl [ \vert F^E_L (q) \vert ^2  + \vert F^M_L (q) \vert ^2 
\bigr ] \Bigr \},  \label{pwba} 
\end{eqnarray}
\begin{eqnarray}
V_{\ell} = { {\rm cos}^2 (\theta /2) \over 4 \, {\rm sin}^4 
(\theta /2)}, \quad 
V_t = { 1 +  {\rm sin}^2 (\theta /2) \over 8 \, {\rm sin}^4 
(\theta /2)},  \nonumber \\ 
V_T = V_t/V_{\ell} = {1 \over 2} + {\rm tg}^2 \left ( {\theta /2}  
 \right ). \label{vlt}
\end{eqnarray}
$Z$ and $e$ are the nuclear proton number and the unit charge,
respectively. Note that the squared form factors  of
\cite{uber,heis} have to be multiplied by $4 \pi /Z^2$ in order to 
obtain our definitions \cite{fanota} for the form factors $\vert
F_L(q) \vert ^2$ (\ref{pwba}), coinciding with those of \cite{bar63}.
Only the term $F^M_1(q)$  survives in (\ref{pwba}) in the case of
$M1$ transitions, while the $E2$ excitations are specified by
$F^C_2(q)$ and $F^E_2(q)$. The PWBA cross section (\ref{pwba}) is
obtained in the relativistic limit (neglected electron rest mass) and
assuming a small excitation energy in comparison with the momentum
transfer, i.e., $E_x \ll q\hbar c$.

It follows from (\ref{vlt}) that the transverse contribution to the
$E2$ cross section is enhanced at $\theta = 165 ^{\circ}$ by a
kinematical factor (\ref{vtrat},\ref{vlt}) $V_T = 58$  with respect
to the longitudinal contribution. On the other hand, it is seen from
Fig.\ \ref{fig1} that the amplitude of the longitudinal transition
density  $\rho _2 ^{\nu} (r)$ is one order of magnitude larger than
the transverse densities ${\cal J} _{2L^{\prime}} ^{\nu} (r), \>
L^{\prime} = 1,3$. The squares of the corresponding form factors will
differ roughly by a factor of 100, since the form factors are Fourier
transforms of the transition densities. Thus, kinematical and
dynamical factors act in opposite directions in backward scattering
and almost compensate each other for $\theta = 165^{\circ}$.

\section{Qualitative analysis of backward scattering}

More definite conclusions can be drawn from Fig.\ \ref{fig7}, where
the PWBA cross section of the $I^{\pi}K=2^+1$ excitation in
$^{156}$Gd is plotted for $\theta = 165^{\circ}, \; 175^{\circ}, \;
178.5^{\circ}$, and $179.5^{\circ}$. The total $E2$ cross section
(continuous curve) is the sum of the longitudinal $C2$ (dashed curve)
and transverse $E2$ (dotted curve) cross sections, as given by
(\ref{pwba}).  For small momentum transfer, $q \approx 0.3$
fm$^{-1}$, the total $E2$ cross section at 165$^{\circ}$ is one order
of magnitude smaller than the corresponding PWBA $M1$ cross section,
shown by a dot-dashed curve in Fig.\ \ref{fig6}.  Thus, the $M1$
dominance over $E2$ at small $q$ is preserved in PWBA, although it is
more pronounced in DWBA where the difference is two orders of
magnitude at smaller $q$, as seen from the top plot of Fig.\
\ref{fig4}.

It is seen from Fig.\ \ref{fig7} that the Coulomb cross section $C2$ 
(dashed curve) provides the main contribution to the $E2$ cross
section at $\theta = 165^{\circ}$ up to a high momentum transfer $q <
1.4$ fm$^{-1}$. The transverse $E2$ cross section remains constant
in all the four plots because the transverse factor $V_t$
(\ref{vlt}) has already reached its maximal value at $\theta =
165^{\circ}$ and does not increase any more appreciably with
$\theta$. In contrast, the longitudinal $C2$ cross section (dashed
curve) is reduced further when going to larger backward angles since
$V_{\ell}$ decreases rapidly. As seen from Fig.\ \ref{fig7}, the
longitudinal and transverse cross sections become comparable at 
$175^{\circ}$. 

The total $E2$ cross section originates exclusively from the
transverse current at $178.5^{\circ}$ in the whole $q$-range studied.
As discussed in Sect.\ II, this is the largest effective angle
reachable in experiments with backward $(e,e^{\prime})$ scattering
\cite{petra}. Further increase in $\theta$ (179.5$^{\circ}$) does not
produce any change in the $E2$ cross section. The transverse
diffraction minima occur at higher $q$-values than the longitudinal
ones because the transverse transition densities ${\cal J}
_{2L^{\prime}} ^{\nu} (r), \> L^{\prime} = 1,3$ are peaked at a
smaller radial distance than the charge transition density $\rho _2
^{\nu} (r)$, as seen from Fig.\ \ref{fig1}. 

The above qualitative analysis in PWBA indicates that the important
$E2$ contributions to the experimental $M1$ cross section at
intermediate transferred momenta and $\theta = 165^{\circ}$, seen in
Figs.\ \ref{fig4},\ref{fig5}, originate mainly from the Coulomb
excitation. Since it is already strongly damped at $\theta =
178.5^{\circ}$, as seen from Fig.\ \ref{fig7}, we plot the DWBA cross
sections for this angle in Fig.\ \ref{fig8}.  It is seen from the
latter figure that the summed cross section (continuous curve)
originates mainly from the $M1$ excitation (dashed curve) for 
$q_{\rm eff} < 1.2$ fm$^{-1}$. Since both $M1$ and $E2$ cross
sections are transverse at this angle, the difference in magnitude
arises mainly from the respective transition densities, shown in
Fig.\ \ref{fig1}.  The two cross sections are of comparable magnitude
for $1.2 < q_{\rm eff} < 2$ fm$^{-1}$. One has to expect an $E2$
dominance for larger momentum transfer (not shown in Fig.\
\ref{fig8}) having in view that lower multipolarities decrease faster
for larger $q$ \cite{tassie}, as discussed in Sect.\ II. 

Therefore, the analysis of the interplay between $M1$ and $E2$
$(e,e^{\prime})$ responses of the low-lying orbital 1$^+$ excitations
leads to the following conclusion.  Even at large backward angles,
where the strong longitudinal $C2$ response is switched off by the
kinematics, the usually expected $M1$ dominance at low $q$ is no more
preserved for larger momentum transfer, $q_{\rm eff} > 1.2$
fm$^{-1}$.

We learned recently \cite{elvira} that the r\^ole of longitudinal
contributions at $\theta = 180 ^{\circ}$ was studied earlier in
even-odd nuclei \cite{mospru}. Their importance was demonstrated for
the ground-state rotational band of $^{181}$Ta. The multipole mixing
is more important in even-odd nuclei since their ground state has a
large angular momentum (spin) and many multipolarities are allowed by
the selection rules. In contrast, even-even nuclei have a zero spin
in the ground state and no multipole mixing occurs in transitions to
the ground state. Thus, the effect we are studying has a different
origin: it is entirely due to the large moment of inertia of heavy
nuclei, as discussed in Sect.\ II.

The DWBA $C2$ cross sections were calculated in \cite{mospru} by
including the Coulomb and neglecting the transverse $E2$
contributions.  The latter, which are strongly enhanced by the
transverse kinematical factor at 180$^{\circ}$, were calculated
separately in PWBA for comparison. They are smaller than the Coulomb
DWBA cross sections but, according to the above analysis, they could
become larger than the longitudinal contribution if calculated in
DWBA. As noted in \cite{mospru}, more precise conclusions could be
drawn after including both longitudinal and transverse contributions
in DWBA calculations, where they interfere in contrast to the
coherent PWBA sum (\ref{pwba}). The transitions within a rotational
band, studied in \cite{mospru}, are characterized by strong Coulomb
excitations because the intrinsic wave function does not change in
the final state and the matrix element of the charge density operator
is large.  In our case of intrinsic excitations with $K^{\pi}=1^+$
the Coulomb excitation is due to the predominantly orbital character 
of the low-lying $M1$ transitions, involving a large non spin-flip
component in the RPA wave function.

The longitudinal kinematical factor $V_{\ell}$ (\ref{vlt}) vanishes
for $\theta = 180 ^{\circ}$, so that a longitudinal contribution can
arise only if the electron rest mass, neglected in (\ref{pwba}),
(\ref{vlt}), is taken into account. The diverging ratio $V_T$
(\ref{vlt}) becomes finite in this case \cite{mospru,theis}:
\begin{equation}
V_T (180 ^{\circ}) \approx \Bigl ( {q\hbar c \over 2 m_e c^2} \Bigr )
^2 \approx \Bigl ( {E_i \over m_e c^2 } \Bigr )^2, \label{vt180}
\end{equation}
where $m_e c^2 = 0.511$ MeV is the electron rest mass.  The
latter relation in (\ref{vt180}) is valid for $E_x \ll E_i$.  It is
seen from (\ref{vt180}) that the longitudinal kinematical factor will
be comparable in magnitude with the transverse one ($V_T \le 1$) only
for unrealistically small momentum transfer $q \le 0.005$ fm$^{-1}$
or incident electron energy $E_i \le 2$ MeV.

A large Coulomb $C2$ cross section was found however in \cite{mospru}
for $\theta = 180 ^{\circ}$ and $q_{\rm eff} < 0.7$ fm$^{-1}$. This
can be understood in PWBA by assuming a transverse electric
excitation of convection type and using the "virtual photon method"
\cite{bar63} for small $q$ to obtain a qualitative estimate for the
relative $CL$ and $EL$ cross sections, given in Table XXIV of
\cite{uber}:
\begin{equation}
{ d\sigma (CL) \over d \sigma (EL)_{\rm conv} } \approx 
\Bigl ( {q \hbar c \over E_x } \Bigr )^2 { 1 \over V_T} \approx 
\Bigl ( { 2 m_ec^2 \over E_x } \Bigr )^2, 
\label{sce}
\end{equation}
where also $V_T (180 ^{\circ})$ from (\ref{vt180}) was used. Note
that the wrong leading $EL$ term $q/M$ was chosen in \cite{willey}
not because it is spin-flip, as stated in \cite{uber}, but because
it is smaller than the correct leading term $E_x/q$ for low $q$.  In
contrast to the severe restrictions (\ref{vt180}), it is seen from
(\ref{sce}) that for any multipolarity $L$ the Coulomb cross section
will be larger than the transverse electric one for excitation
energies $E_x \le 1$ MeV.  This is indeed the case in \cite{mospru},
where the studied first two members of the ground state rotational
band have very small energies, 0.14 and 0.3 MeV.

The situation is more unfavourable in our case, where the considered
1$^+$ excitations from Table\ \ref{tabrpa} have $E_x \approx 3$ MeV.
In this case we have still a dominant Coulomb contribution at
165$^{\circ}$ for $q < 1.2$ fm$^{-1}$, but for $\theta > 178
^{\circ}$ the $C2$ cross section is negligible in comparison with the
transverse $E2$ cross section, as seen from Fig.\ \ref{fig7}. This
result agrees with the qualitative PWBA prediction (\ref{sce}) for a
difference of one order of magnitude at $180^{\circ}$. Therefore, our
results differ qualitatively from those for the rotational
excitations of even-odd nuclei \cite{mospru}, where the attention was
focused on a fully backward angle, $\theta = 180 ^{\circ}$, so that
the main effect found there  (large $C2$ contributions)  is due to
the small but non-negligible electron rest mass (\ref{vt180}),
(\ref{sce}). This is not the case with the 1$^+$ excitations in
even-even nuclei, studied in the present work, where the Coulomb
contribution becomes negligible already before reaching a full
backward angle.

Thus, in our case of low-lying orbital 1$^+$ excitations the
interplay of transverse electric and magnetic contributions is of
prime importance for $\theta > 178 ^{\circ}$. We borrow the PWBA
estimates in the low-$q$ limit from Table XXIV of \cite{uber}:
\begin{equation}
{ d\sigma (M,L-1) \over d \sigma (EL)_{\rm conv} } \approx 
\Bigl ( {\hbar ^2 c^2 \over E_x R^2 Mc^2 } \Bigr )^2 \approx
E_x^{-2}, \label{sme}
\end{equation}
where $M$ is the nucleon mass.  A nuclear radius $R=6.5$ fm was
assumed for the rare-earth region to obtain the latter relationship
in (\ref{sme}), where the excitation energy $E_x$ must be supplied in
MeV. The spin contribution to the transverse electric cross section
is negligible with respect to the convection contribution at low $q$
and the ratio between the $M1$ and $E2$ cross sections is determined
by (\ref{sme}). This relationship  should be valid for any scattering
angle. It applies to both the convection and spin parts of the
magnetic excitation $(M,L-1)$.

It is seen from (\ref{sme}) that a dominant $M1$ excitation is
predicted in PWBA for small rotational energies, e.g. $E_x \le 0.3$
MeV \cite{mospru}, but not in our case where $E_x \approx 3$ MeV 
(Table\ \ref{tabrpa}). The PWBA expectation for a dominant $E2$
transition, which should be valid at least for low $q$, is not
supported by our microscopic results, shown in Fig.\ \ref{fig8}: the
$M1$ cross section (dashed curve) is one order of magnitude larger
than the $E2$ response (dotted curve) for $q_{\rm eff} < 1.2$
fm$^{-1}$.

The main features of the competing $E2$ and $M1$ responses to
inelastic electron scattering, found by the qualitative analysis in
this section, are common for all the low-lying 1$^+$ excitations
from Table\ \ref{tabrpa}, studied in this work. Each of them is the
strongest orbital $M1$ excitation in the respective nucleus and has
the character of a weakly collective scissors mode
\cite{nofa88,nfsciss}. 

\section{Summary and conclusions}

We have studied the competition between electric and magnetic
excitations with $K^{\pi} =1^+$ in response to inelastic electron
scattering from heavy deformed nuclei. This study, whose initial 
results were reported briefly in \cite{nfd94,dnf95}, is, to our
knowledge, the first theoretical investigation of possible electric
contributions to $M1$ cross sections  of even-even nuclei,
measured experimentally at backward scattering angle. The topic is of
some general interest, because the backward electron stattering has
established itself since three decades as one of the main tools for
the experimental study of nuclear magnetic dipole excitations due to
its particular sensitivity towards the magnetic response at backward
angles \cite{pet,bar63,bar62,fagg,uber,ramfagg}. 

Electric contributions have been usually neglected at backward
angles, apart from a theoretical indication \cite{mospru} for
a possible Coulomb excitation in the ground state rotational band of
heavy even-odd nuclei on the example of $^{181}$Ta. The multipole
mixing is allowed in even-odd nuclei since they have a non-vanishing
angular momentum in the ground state. In contrast, the mixing is
forbidden by selection rules for transitions to the ground state of
even-even nuclei, which has a zero spin.

However, a rotational band develops in deformed nuclei on the top of
each intrinsic excitation with $K^{\pi}=1^+$. The  $M1$ transition
with $I^{\pi}K=1^+1$ is accompanied by an $E2$ transition to the
first rotational member with $I^{\pi}K=2^+1$ of this band. The energy
separation between the two excitations is small in heavy nuclei with
a large moment of inertia and comparable with the experimental energy
resolution in electron scattering. Such a mechanism for multipole
mixing was not considered previously, although it gains in importance
through the systematic occurrence of accompanying $E2$ transitions. It
stands in contrast to the usual low probability for the random
presence of some other intrinsic $E2$ excitation in the vicinity of
the studied $M1$ transition. 

The accompanying $E2$ transition was identified experimentally only
in $^{156}$Gd as yet \cite{bohle85}.  We were motivated by this work
in the search for a possible explanation for the systematic
deviations of the theoretical $M1$ form factors from experiment at
higher momentum transfer, occurring only in heavy nuclei, but absent
in light ones \cite{nofalip} even up to higher transferred momentum.

We describe the 1$^+$ excitations in heavy nuclei within QRPA using a
deformed Woods-Saxon potential, BCS pairing, and separable residual
interactions. Our formalism  was able to provide a good description
of 1$^+$ excitations (see the review \cite{noj95}) in Titanium
\cite{nofalip,fanota}, rare-earth \cite{fanos,nfsciss,sarri} and
actinide \cite{noj93} nuclei, studied in $(e,e^{\prime})$, $(\gamma ,
\gamma ^{\prime})$, and $(p,p^{\prime})$ experiments, e.g.,
\cite{bohle85,bohle86,bohle84,hart89,dja85,pitz89,boh84}, reviewed in
\cite{achimrev,achim90}. 

Using this formalism, we have studied here the competing $E2$ and
$M1$ excitations in all the rare-earth nuclei with experimentally
measured $(e,e^{\prime})$ cross sections. These are the strongest
$M1$ excitations in each  nucleus. They have an orbital nature and
correspond to a weakly collective low-energy scissors mode
\cite{nofa88,nfsciss}. The theoretical energies, $B(M1)$ values, and
orbit-to-spin ratios are in agreement with experiment. The only
available experimental $B(E2)$ value \cite{bohle85} is also
reproduced. The necessary transition densities are obtained
microscopically from the QRPA wave functions within our formalism,
developed in a cylindrical basis. The $(e,e^{\prime})$ cross sections
are calculated in DWBA from the transition densities. 

The comparison with the experimental $(e,e^{\prime})$ cross sections,
measured at scattering angle $\theta = 165 ^{\circ}$, shows that the
previous systematic deviations of the theoretical $M1$ cross
sections from experiment at higher momentum transfer $q$ are removed
by taking the $E2$ response into account. At low $q$ the theoretical
$E2$ cross section is two orders of magnitude smaller than the
corresponding $M1$ cross section. Therefore, the $E2$ response does
not introduce any corrections to the $B(M1)$ values derived by 
extrapolating the experimental cross sections to the photon point
(minimal $q$ value).

The $E2$ and $M1$ cross sections become comparable in magnitude in
the region $0.7 < q_{\rm eff} < 0.8$ fm$^{-1}$ ($0.5 < q < 0.6$
fm$^{-1}, \; 50 \le E_i \le 60$ MeV). The accompanying $E2$
response provides an important contribution to the measured cross
sections for $0.6 < q_{\rm eff} < 0.9$ fm$^{-1} \> (0.4 < q < 0.7$
fm$^{-1}, \; 40 \le E_i \le 70$ MeV).  The sum of the $E2$ and $M1$
theoretical cross sections agrees quite well with the available
experimental data. The qualitative behaviour, the improvement, and
the achieved agreement with experiment are manifested in a similar
way in all the excitations studied and exhibit a well pronounced
systematic character. This allows us to conclude that the $E2$
response of low-lying orbital 1$^+$ excitations to inelastic electron
scattering plays an important r\^ole in heavy deformed nuclei for
higher momentum transfer even at backward scattering angles.

The PWBA analysis of the cross sections has shown that the $E2$
excitation is dominated by the Coulomb $C2$ response at  
$\theta = 165 ^{\circ}$ and this qualitative feature is preserved for
$\theta \le 175 ^{\circ}$. In contrast, the longitudinal $C2$
response is already negligible for the largest experimentally
reachable effective backward angle $\theta = 178.5 ^{\circ}$
\cite{petra}. It is simply damped by the small longitudinal kinematic
factor. The interplay between magnetic and electric responses is
determined at this angle (and beyond it until 180$^{\circ}$) by the
competing transverse magnetic and electric responses. Both of them
receive the main contribution from the proton convection current. 

The transverse $M1$ response is still dominant at $178.5 ^{\circ}$
for $q_{\rm eff} < 1.2$ fm$^{-1}$, corresponding to incident electron
energy $E_i \le 100$ MeV. The two responses are comparable beyond
this energy and the $E2$ cross section is expected to be even the
dominant one for $E_i > 200$ MeV. Therefore, the commonly expected
$M1$ dominance in backward scattering does not take place at high
momentum transfer even when the full backward angle is reached. This
is true at least for the orbital $M1$ excitations studied in the
present work.

\acknowledgments

 Thanks are due to Henk P. Blok and Jochen Heisenberg for providing
us with their DWBA code and useful communications on the underlying
formalism. This work is supported by the Deutsche
Forschungsgemeinschaft (DFG) under contract Fa67.

\appendix
 
\section{Cylindrical basis}

   A large number of formulas for multipole transition densities has
been derived in the past using non-trivial relationships from the
differential tensor calculus, see, e.g., \cite{uber} and
references therein. They are very useful for microscopic
calculations, but almost all of them are formulated in a spherical
basis, whose properties are essential for obtaining the final
results. The spherical basis is not suitable for deformed
nuclei. We follow the appoach of our former work \cite{fanota} and
derive the necessary expressions for deformed basis in cylindrical
coordinates {\bf r} = ($\rho , z, \varphi$), where $\rho$ and
$\varphi$ are the azymuth radius and angle, and $z$ is the symmetry
axis of the nucleus. 

   We are using a deformed, axially symmetric Woods-Saxon potential 
\cite{dam69}, whose eigenfunctions are expanded in the basis of the
eigenfunctions of the deformed, axially symmetric harmonic oscillator
\begin{eqnarray}
\Psi _j ({\bf r}) = 
\Psi (n_{\rho},n_z,\Lambda,\Sigma,{\bf r}) \nonumber \\
= \psi (n_{\rho}, \vert \Lambda \vert , \rho ) \, \psi ( n_z , z) \, 
\psi (\Lambda , \varphi ) \, \chi (\Sigma ),   
\label{basis} \end{eqnarray}
where $j = (n_{\rho}, n_z, \Lambda , \Sigma )$ denotes a set of four 
quantum numbers fully specifying the wave function. $ \chi (\Sigma )$
is a two-component spinor with projection $ \Sigma = \pm 1/2$ of the
spin {\bf S} on the quantization z-axis. The principal quantum number
${\cal N}$, parity $\pi$, projection $\Lambda$ of the orbital angular
momentum {\bf L} and projection $K$ of the total angular momentum
{\bf J} = {\bf L} + {\bf S} on the nuclear symmetry axis $z$ are
related in the following way:
\begin{eqnarray}
{\cal N} = n_z + 2 n_{\rho} + \vert \Lambda \vert , \nonumber \\
K = \Lambda + \Sigma, \quad \pi = (-1) ^{\cal N}.  
\label{qnum} \end{eqnarray}
 
   The remaining functions in (\ref{basis}) are expressed
\cite{volkof} through the Hermite and associate Laguerre  polynomials
H$_{n_z}$,  L$_{n_{\rho}} ^{\vert \Lambda \vert}$:
\begin{eqnarray}
 \psi (n_{\rho}, \vert \Lambda \vert , \rho )  & = & \psi _j
^{\Lambda} (\rho ) = {\rm N}_{n_{\rho}} ^{\vert \Lambda \vert}  
\sqrt{2}  {\rm R}_{\perp}  \eta^{\vert \Lambda \vert \over 2}  
{\rm exp} (-{\eta \over 2})  {\rm L}_{n_{\rho}} ^{\vert \Lambda
\vert} (\eta),  \nonumber \\
\psi ( n_z , z)  & = & \psi _j (z) = {\rm N}_{n_z} \, 
\sqrt{{\rm R}_z} \, {\rm exp} (-{\xi^2 \over 2}) \, {\rm H}_{n_z} 
(\xi),  \nonumber \\
\psi (\Lambda , \varphi ) & = & \frac{1}{\sqrt{2\pi}} \, {\rm exp} 
(i \Lambda \phi ),  
\label{psibas} \end{eqnarray} 
where shorter notations are introduced for the wave functions with
$j$ denoting either $n_{\rho}$ or $n_z$, to remind that both of them
belong to the same set of four quantum numbers specifying the
eigenfunction (\ref{basis}). The norms N and the coordinate scaling
factors R are given by
\begin{eqnarray}
{\rm N}_{n_{\rho}} ^{\vert \Lambda \vert } = 
\sqrt{ n_{\rho} ! / (n_{\rho} + \vert \Lambda \vert )! } \, , \;
 {\rm N}_{n_z} = (\sqrt{\pi} \, 2^{n_z} \, n_z !) ^{-\frac{1}{2}}, 
\nonumber \\ 
\eta = \rho ^2  {\rm R}_{\perp}^2 , \quad \xi = z {\rm R}_z , 
\nonumber \\ 
 {\rm R}_{\perp} = \sqrt{M \omega _{\perp}/ \hbar} , \quad 
 {\rm R}_z = \sqrt{M \omega _z / \hbar} . 
\label{nrfact} \end{eqnarray}
M is the nucleon mass and $\omega _{\perp}, \; \omega_z$ are the two
frequences of the axially symmetric harmonic oscillator. The
time-reversed basis wave functions (\ref{basis}) have the form
\begin{equation}
{\tilde \Psi} _j ({\bf r}) = \hat{T} \Psi _j ({\bf r}) = (-1) 
^{1/2 - \Sigma} \Psi (n_{\rho},n_z, - \Lambda, - \Sigma,{\bf r}). 
\label{tpsi} \end{equation} 

   Let us introduce short notations $D$ for the derivatives of
(\ref{psibas}) with respect to $\rho$ and $z$:
\begin{eqnarray*}
{\partial \over \partial \rho} \psi (n_{\rho}, \vert \Lambda 
\vert , \rho ) = D_j ^{\Lambda} (\rho ) + {\vert \Lambda \vert \over 
\rho} \psi (n_{\rho}, \vert \Lambda \vert , \rho ), \\
 D_j  ^{\Lambda} (\rho ) = - {\rm R}_{\perp} \Bigl [ \sqrt{n_{\rho}} 
 \, \psi (n_{\rho} -1, \vert \Lambda \vert +1, \rho ) \\ 
 + \sqrt{n_{\rho} + \vert \Lambda \vert +1} \, 
\psi (n_{\rho}, \vert \Lambda \vert +1, \rho ) \Bigr ], 
\end{eqnarray*}
\begin{eqnarray}  
D_j (z) &=& {\partial \over \partial z} \psi (n_z, z)  \nonumber \\
& = & {{\rm R}_z \over \sqrt{2}} \Bigl [ \sqrt{n_z} \, 
\psi (n_z -1, z) -  \sqrt{n_z +1} \, \psi (n_z +1, z) \Bigr ],
\nonumber \\ 
& & {\partial \over \partial \varphi} \psi (\Lambda , \varphi ) =
i \Lambda \psi (\Lambda , \varphi ). 
\label{deriv} \end{eqnarray}

   The eigenfunctions $\Phi_i$ of the axially symmetric Woods-Saxon
potential \cite{dam69} are linear combinations of (\ref{basis}) with
definite projection $K$ and parity $\pi$. Different functions are
labelled with $i$, because there are many orthogonal functions with
the same $K^{\pi}$. For $K = K_i \ge 1/2$:
\FL \begin{eqnarray}
\Phi_i &(& K ^{\pi}, {\bf r} ) = \sum _j C^i_j \Psi_j (n_{\rho},
n_z,\Lambda,\Sigma = \textstyle{1 \over 2}, K, {\bf r})
\nonumber \\
&& + \sum _{j^{\prime}} C^i_{j^{\prime}} \Psi_{j^{\prime}}
(n^{\prime} _{\rho}, n^{\prime}_z, \Lambda^{\prime} = \Lambda +1, 
\Sigma^{\prime} = - \textstyle{1 \over 2}, K, {\bf r}), \nonumber \\
 && \hspace*{4 truecm} \Lambda = K - 1/2, \label{wswf}
\end{eqnarray}
where $C^i_j, \; j = (n_{\rho}, n_z)$, are expansion coefficients
found after numerical diagonalization of the hamiltonian in some
truncated basis with ${\cal N} \le {\cal N}_{\rm max.}$. The
expansion (\ref{wswf}) is separated explicitly in two terms with
constant $(\Lambda , \Sigma)$, because this property will be used
below in the analytical derivation of expressions for transition
densities. The time-reversed state corresponding to (\ref{wswf}) is
easily obtained from (\ref{tpsi}):
\FL \begin{eqnarray}
{\tilde \Phi} _i &(& K ^{\pi}, {\bf r}) = \sum _j C^i_j  
\Psi _j (n_{\rho},n_z, - \Lambda,\Sigma = - \textstyle{1 \over 2}, 
-K, {\bf r}) 
\nonumber \\
&& - \sum  _{j^{\prime}} C^i_{j^{\prime}} \Psi_{j^{\prime}} 
(n^{\prime}_{\rho}, n^{\prime}_z, - \Lambda -1, \Sigma^{\prime} = 
\textstyle{1 \over 2}, -K, {\bf r}). \label{twswf} 
\end{eqnarray}
We describe intrinsic excitations with $\Delta K = \vert K_f - K_i 
\vert =1$.  Thus, time-reversed states appear only in matrix elements
between initial and final states with $K_i = -1/2, \; K_f = 1/2$ and
$\Lambda = 0$. It is explained in Sect. D how they can be obtained
from the expressions (\ref{jcl}), (\ref{jsl}) by some phase changes. 

\section{Symmetries of operators}

The charge and current density operators (\ref{rhop},\ref{jtop}) have
the following properties under time-reversal $\hat{T}$ and hermitian
conjugation: 
\begin{eqnarray*}
\hat{\rho} ^{\dag} ({\bf r}) = \hat{\rho} ({\bf r}), \quad 
\hat{T} \hat{\rho} ({\bf r}) \hat{T}^{-1} = \hat{\rho} ({\bf r}), 
\\
\hat{\bf J} ^{\dag} ({\bf r}) = \hat{\bf J} ({\bf r}), \quad 
\hat{J} ^{\dag}_{\mu} ({\bf r}) = (-1) ^{\mu} \hat{J} _{- \mu} 
({\bf r}),  \\
\hat{T} \hat{\bf J} ({\bf r}) \hat{T}^{-1} = - \hat{\bf J} ({\bf r}), 
\quad 
\hat{T} \hat{J}_{\mu} ({\bf r}) \hat{T}^{-1} = (-1) ^{1+ \mu} \hat{J} 
_{- \mu} ({\bf r}), 
\end{eqnarray*}
\begin{equation}
\hat{J}_0 = \hat{J}_z, \quad \hat{J}_{\mu} = - {\mu \over \sqrt{2}}
\left ( \hat{J}_x + \mu \hat{J}_y \right ), \quad \mu = \pm 1, 
\label{sdop} \end{equation}
where $\hat{J}_{\mu}, \; \mu =0, \pm 1$ are the cyclic (tensorial)
components of the vector. It follows from (\ref{sdop}) that the
multipole density operators (\ref{mdop}) have the properties 
\begin{eqnarray} 
\hat{\rho} ^{\dag} _{LM} (r) = (-1)^{M} \hat{\rho} _{L,-M}(r), 
\nonumber \\ 
\hat{T} \hat{\rho} _{LM} (r) \hat{T}^{-1} = (-1)^{M} \hat{\rho} 
_{L,-M}(r), \nonumber \\
\hat{\cal J} ^{\dag}_{LL^{\prime}M} (r) = (-1) ^{L + L^{\prime}
+ M +1} \hat{\cal J} _{LL^{\prime},-M} (r), \nonumber \\
\hat{T} \hat{\cal J} _{LL^{\prime}M} (r) \hat{T}^{-1} = 
(-1) ^{L + L^{\prime} + M} \hat{\cal J} _{LL^{\prime},-M} (r). 
\label{smdop}
\end{eqnarray}

  The $m$-symmetrized operators (\ref{mop}) change only a phase
under hermitian and time conjugation.  Let us denote with c$^{\rm H}$,
c$^{\rm T}$ the corresponding phases of such an operator $\hat{B}
(m)$:
\begin{equation}
\hat{B} ^{\dag} (m) = {\rm c}^{\rm H} \hat{B} (m), \quad 
\hat{T} \hat{B} (m) \hat{T}^{-1} = {\rm c}^{\rm T} \hat{B} (m). 
\label{phase} \end{equation}
It follows from (\ref{smdop}) and corresponding relationships for
$\hat{Q} (m), \; \hat{J} (m)$ that the  $m$-symmetrized operators
(\ref{mop}) have the properties listed in Table \ref{tabphas}.

The spin and orbital angular momentum operators are symmetrized in
the same way as $\hat{J}(m)$ in (\ref{mop}).  It is seen from
Table \ref{tabphas} that all transition  operators have real matrix
elements in our cylindrical basis
(\ref{basis}),(\ref{wswf}),(\ref{twswf}), apart from the current
density operators $\hat{\cal J} _{LL^{\prime}} (m,r)$, whose matrix
elements are imaginary. Operators with $m = +1 (-1)$ are hermitian
(antihermitian), apart from the angular momentum and spin operators
in the last column of Table \ref{tabphas}, which have the opposite
property.  

  All the operators acquire a phase of the type c$^{\rm C} {\rm c}
^{\rm T} = \pm m$. This phase can always be chosen to be $+m$ by
redefining $\hat{B} (m) \to \hat{B} (-m)$, if necessary. The
convention
\begin{equation}
 {\rm c}^{\rm C} {\rm c} ^{\rm T} = +m  \label{ctphas} 
\end{equation}
has already been taken in the definitions (\ref{mop}) into account,
as seen from Table \ref{tabphas}.  It can be shown that, after
adhering to (\ref{ctphas}) and neglecting in QRPA the scattering
terms $\alpha ^{\dag} \alpha, \; \alpha \alpha ^{\dag}$, any operator
with the properties (\ref{phase}) has the following quasiboson
representation: 
\begin{eqnarray}
\hat{B}_{\sigma} (m) = \sum _{i,f > 0} \Bigl \{ b_{\sigma} (fi,m) 
\nonumber \\ 
\times [ A^{\dag} (if,m) + \sigma \> m A(if,m) ] \nonumber \\
 + b_{\sigma} (\tilde{f}i,m) [ \tilde{A}^{\dag} (if,m) + \sigma \> 
m \tilde{A}(if,m) ] \Bigr \}, \label{bosrep} 
\end{eqnarray}
\begin{eqnarray} 
b_{\sigma} (fi,m) = -{1 \over \sqrt{2}} (u_f v_i + \sigma \> u_i v_f)  
\langle f \vert \hat{B}_{\sigma} (m) \vert i \rangle , \nonumber \\ 
b_{\sigma} (\tilde{f}i,m) = {1 \over \sqrt{2}}  (u_f v_i + \sigma \> 
u_i v_f) \langle \tilde{f} \vert \hat{B}_{\sigma} (m) \vert i 
\rangle ,  \label{qpme} \\
\sigma = {\rm c}^{\rm TH} = {\rm c} ^{\rm T} {\rm c} ^{\rm H} = 
\pm 1,  \label{sigma} 
\end{eqnarray}
where $u,v$ are the BCS occupation probabilities and 
the indices $(i,f)$ denote the Woods-Saxon eigenfunctions
(\ref{wswf}),(\ref{twswf}), used to calculate the single-particle
matrix elements in (\ref{bosrep}). The summation runs only over
states with $K > 0$  because time-reversed states (with $K < 0$) are
treated explicitly in (\ref{bosrep}). 

   The q.p. basis is provided by the $m$-symmetrized 2q.p. creation
and annihilation operators \cite{nofa88}, appearing in
(\ref{bosrep}),
\begin{eqnarray}
 A^{\dag} (if,m) = {1 \over \sqrt{2}} ( \alpha ^{\dag} _{\tilde{i}} 
\alpha ^{\dag} _f - m \, \alpha ^{\dag} _i \alpha ^{\dag} _{\tilde{f}} 
), \nonumber \\
  \tilde{A}^{\dag} (if,m)  = {1 \over \sqrt{2}} 
(  \alpha ^{\dag} _{\tilde{f}}  \alpha ^{\dag} _{\tilde{i}} 
 + m \, \alpha ^{\dag} _f \alpha ^{\dag} _i ). \label{aqp} 
 \end{eqnarray} 
Basis operators (\ref{aqp}) with different $m$-values commute and
determine the same property of the quasiboson operators
(\ref{bosrep}), 
\begin{eqnarray}
 [ A(if,m), \;  A^{\dag} (if,m^{\prime}) ] = 0, \nonumber \\
\,  [ \hat{B}_{\sigma} (m), \;  \hat{B}^{\dag} _{\sigma} (m) ] = 0, 
\nonumber \\
\,  [ \hat{B}_{\sigma ^{\prime}} (m^{\prime}), \;  \hat{B} _{\sigma} 
(m) ] = 0,  \quad  \text{ for } m^{\prime} \ne m, 
\nonumber \\
\,  [ \hat{B}_{\sigma} (m), \;  \hat{B} _{\sigma^{\prime}} (m) ]
\ne  0,  \quad  \text{ for } \sigma^{\prime} \ne \sigma .   
\label{comut} 
\end{eqnarray}
Creation and annihilation operators with the same $\sigma , \, m$
commute as well, because the hermitian conjugation produces only a
phase (\ref{phase}). This property is a direct consequence from the
fact that the QRPA vacuum $\Gamma _{\mu} (m) \vert \rangle$
(\ref{phon}) is not a vacuum for the quasiboson operators
(\ref{bosrep}), i.e., $\hat{B} _{\sigma} (m) \vert \rangle \ne 0$.  
Only operators with the same $m$ and different $\sigma$ (longitudinal
and transverse) have a non-vanishing commutator, reflected in the
last relation (\ref{comut}). 

  It is seen from Table \ref{tabphas} that the quadrupole and charge
density operators are characterized by $\sigma = 1$ (longitudinal
type), while the spin, angular momentum, and current density
operators, corresponding to axial vectors in the $(x,y)$-plane,  have
$\sigma = -1$ (transverse type). This fundamental property can not be
changed by redefining the $m$-symmetrization in (\ref{mop}). The
q.p. matrix elements of $\hat{Q}(m)$ and $\hat{J}(m)$ are given in
\cite{nofa88}. Note that we are using here a different definition of
$\hat{J}(m)$  \cite{nofa93,nofa94}.  The q.p. matrix elements
(\ref{qpme}) of the density operators (\ref{mop}) have the form:
\begin{eqnarray*}
\rho _2 (fi,m,r) = - { 1 \over \sqrt{2}} U(f,i) \langle f \vert
\hat{\rho} _2 (m,r) \vert i \rangle, \nonumber \\
\rho _2 (\tilde{f}i,m,r) = { 1 \over \sqrt{2}} U(f,i) \langle
\tilde{f} \vert \hat{\rho} _2 (m,r) \vert i \rangle, \nonumber \\
U(f,i) = u_f v_i + u_i v_f, \quad V(f,i) =  u_f v_i - u_i v_f,
\end{eqnarray*}
\begin{eqnarray}
{\cal J} _{LL^{\prime}} (fi,m,r) =  - { 1 \over \sqrt{2}} V(f,i)
\langle f \vert  \hat{\cal J} _{LL^{\prime}} (m,r) \vert i \rangle,
\nonumber \\
{\cal J} _{LL^{\prime}} (\tilde{f}i,m,r) = { 1 \over \sqrt{2}} V(f,i)
\langle \tilde{f} \vert  \hat{\cal J} _{LL^{\prime}} (m,r) \vert i
\rangle. \label{mopqp} 
\end{eqnarray}

  The q.p. matrix elements (\ref{mopqp}) have the same symmetries as
those of the remaining operators (\ref{mop}). Thus, one can write in
general for the q.p. matrix elements of any operator
$\hat{B}_{\sigma} (m)$ (\ref{mop}), among those listed in Table
\ref{tabphas},
\begin{eqnarray}
B(fi,m) & = & m B(if,m) = m B(\tilde{f}\tilde{i},m) = B(\tilde{i}
\tilde{f},m), \nonumber \\
B(\tilde{f}i,m) & = & m B(i\tilde{f},m) \nonumber \\ 
 & = & -m B(f\tilde{i},m) = - B(\tilde{i}f,m), \label{qpsym}  
\end{eqnarray}
where the phase $\sigma$ and the radial dependence of density 
operators are skipped for simplicity. The definitions of
the $m$-symmetrized operators (\ref{mop}) have been specially chosen
to ensure the above uniform properties of their matrix elements,
valid for both longitudinal ($\sigma =1$) and transverse ($\sigma =
-1$) operators.

\section{Multipole transitions}

  The QRPA phonon creation operators \cite{nofa88} are expanded over
the q.p. basis (\ref{aqp}),
\begin{eqnarray}
\Gamma ^{\dag} _{\nu} (m) = {1 \over 2} \sum _{i,f > 0} 
\bigl [ \psi _{\nu} (fi,m) A^{\dag} (if,m) \nonumber \\
- \phi _{\nu} (fi,m) A(if,m) \nonumber \\
 + \lambda _{\nu} (fi,m) \tilde{A}^{\dag} (if,m) - \mu _{\nu} (fi,m)
\tilde{A}(if,m) \bigr ], \label{phon}
\end{eqnarray}
where $\psi, \, \lambda$ are forward- and $\phi, \, \mu$
backward-going RPA amplitudes, found numerically for each excited
state $\nu$ by solving the RPA equations of motion. The
$m$-symmetrized wave function is obtained in the laboratory frame
after projecting the intrinsic excitation $\Gamma ^{\dag} _{\nu} (m)
\vert \rangle$ on a good angular momentum \cite{nofa88}. One obtains
with this wave function the reduced matrix element of any transition
operator $\hat{T}^{\dag}_L$ (irreducible tensor of rang $L$) between
the nuclear ground state and the considered $\nu$-th RPA excitation
with a total angular momentum $I$, its projection $K$ on the nuclear
symmetry axis, and parity $\pi$: 
\FL \begin{eqnarray}
\langle I^{\pi}K= L^+&1,& \nu \parallel  \hat{T}^{\dag}_L 
\parallel 0, \text{g.s.} \rangle \nonumber \\ 
 &=&  {1 \over 2} \sum _{m=\pm 1} \Bigl [ \langle \vert \Gamma _{\nu} 
 (m) \hat{T}^{\dag}_{L1} \vert \rangle \nonumber \\
&+& (-1)^{L+1} m \, \langle \vert \Gamma _{\nu} (m)
\hat{T}^{\dag}_{L,-1} \vert \rangle \Bigr ]. \label{rme}
\end{eqnarray}
The multipolarity $L$ of $\hat{T}^{\dag}_L$ in (\ref{rme}) specifies
the corresponding $m$-symmetrized operators from (\ref{mop}), i.e.,
$L=1$ refers to $\hat{\cal J} _{11} (m,r), \, \hat{J}(m), \,
\hat{L}(m)$, and $\hat{S}(m)$, while $L=2$ means $\hat{\rho}_2 (m,r),
\, \hat{\cal J} _{2L^{\prime}} (m,r)$, or $\hat{Q} (m)$. The
necessary cyclic components $\hat{T} ^{\dag} _{L,\pm 1}$, appearing
in (\ref{rme}), are easily obtained from (\ref{mop}), where all the
operators are taken in the quasiboson representation (\ref{bosrep}):
\begin{eqnarray}
\hat{T}^{\dag}_{LM} = \hat{B}_{\sigma} (m=-\sigma ) + M \, 
\hat{B}_{\sigma} (m=\sigma ), \quad \hat{B} \ne \hat{J}, \nonumber \\
\hat{J}^{\dag}_{1M} = \hat{T}^{\dag}_{1M}/\sqrt{2}, \quad 
\text{ for } \hat{B} (m) = \hat{J} (m).    \label{tensym}
\end{eqnarray}

  The matrix elements in (\ref{rme}), calculated with the help of
(\ref{tensym}), have the form:
\begin{eqnarray}
\langle \Gamma _{\nu} (m) \> \hat{B}_{\sigma} (m,r) \rangle = 
\langle \bigl [ \Gamma _{\nu} (m), \> \hat{B}_{\sigma} (m,r) \bigr ] 
\rangle \nonumber \\ 
= \sum _{i,f > 0} \Bigl [ F^{\nu} _{\sigma} (fi,m) b_{\sigma} 
(fi,m,r) \nonumber \\
+  F^{\nu} _{\sigma} (\tilde{f}i,m) b_{\sigma} (\tilde{f}i,m,r) 
\Bigr ], \label{intme} 
\end{eqnarray}
where $b_{\sigma}(fi,m,r)$ are the q.p. matrix elements of $\hat{B}
_{\sigma} (m,r)$ (\ref{qpme}).  They are given by (\ref{mopqp}) in
the case of transition density operators.  The coefficients $F^{\nu}
_{\sigma} (fi,m)$ are linear combinations of the RPA amplitudes
(\ref{phon}):
\begin{eqnarray}
 F^{\nu} _{\sigma} (fi,m) = \psi _{\nu} (fi,m) + \sigma m \, 
 \phi _{\nu} (fi,m), \nonumber \\ 
 F^{\nu} _{\sigma} (\tilde{f}i,m) = \lambda _{\nu} (fi,m) + \sigma 
 m \, \mu _{\nu} (fi,m).  \label{fsig}
\end{eqnarray}
A different notation, $F_{\nu}(fi,m) \equiv  F^{\nu} _{+1} (fi,m),
\> G_{\nu}(fi,m)$ $\equiv  F^{\nu} _{-1} (fi,m)$, was used in our
previous works, e.g. \cite{nofa88,fanota,fanos}. It can be shown that
the amplitudes (\ref{fsig}) have exactly the same properties as the
q.p.  matrix elements (\ref{qpsym}). The definitions (\ref{mop})
ensure that these properties are the same for the q.p.  matrix
elements of all the operators (\ref{mop}). One can show in this way
that the intrinsic transition matrix elements (\ref{intme}) do not
depend on the signature $m$, i.e.,
\begin{eqnarray}
\langle \Gamma _{\nu} (m=+1) \> \hat{B}_{\sigma} (m=+1,r) \rangle
\nonumber \\
= \langle \Gamma _{\nu} (m=-1) \> \hat{B}_{\sigma} (m=-1,r) \rangle.
\label{intpm} \end{eqnarray}
Using (\ref{tensym}) and (\ref{intpm}), one obtains for the reduced
transition matrix elements of the type (\ref{rme}):
\[ \langle I^{\pi}K=L^+1, \nu \parallel \hat{T} _L \parallel 0, 
\text{g.s.} \rangle = 0, \]
\begin{equation}
\langle L^{\pi}1, \nu \parallel \hat{T}^{\dag}_L  \parallel 0, 
\text{g.s.}  \rangle = 2 \, \langle \Gamma _{\nu} (m) \> 
\hat{B} _{\sigma} (m,r) \rangle,  \label{rmel} 
\end{equation}
where the intrinsic matrix element in the r.h.s of (\ref{rmel}), 
given by (\ref{intme}), (\ref{fsig}), can be calculated either for
$m=1$ or for $m=-1$. It is seen from (\ref{rmel}) that in our
$m$-symmetrized formalism the QRPA ground state is not only a vacuum
for the QRPA annihilation operators (\ref{phon}), $\Gamma _{\nu} (m)
\vert \rangle = 0$, but also for the transition operators $\hat{T}
^{\prime}$ in the laboratory frame. This property is not trivial,
because it is not fulfilled for the corresponding transition
operators $\hat{B}_{\sigma} (m,r)$,  $\hat{B} _{\sigma} ^{\dag}(m,r)$
(\ref{mop}) in the intrinsic frame: both of them have a non-vanishing
action on the QRPA vacuum, as it can be seen from (\ref{intme})
having in view the hermitian properties (\ref{phase}), listed in
Table \ref{tabphas}.

  Half of the terms in microscopic sums of the type (\ref{intme}) are
redundant because of the relationships (\ref{qpsym}). Thus, we make
a transition from $m$-symmetrized (\ref{mop}) to the usual step-up
operators $\hat{J}_+, \, \hat{L}_+, \, \hat{S}_+$ (\ref{mop}) or
tensors with $M=+1$ for the remaining operators (\ref{mop}). Let us 
define with them the following q.p. matrix elements:
\begin{eqnarray}
\rho _{21} (fi,r) = U(f,i) \langle f \vert \hat{\rho} _{21} (r) \vert
i \rangle, \nonumber \\ 
\rho _{21} (f \tilde{i},r) = U(f,i) \langle f \vert \hat{\rho} _{21} 
(r) \vert \tilde{i} \rangle, \nonumber \\
{\cal J} _{LL^{\prime}1} (fi,r) = V(f,i) \langle f \vert \hat{\cal J}
_{LL^{\prime}1} (r) \vert i \rangle, \nonumber \\ 
{\cal J} _{LL^{\prime}1} (f \tilde{i},r) = V(f,i) \langle f \vert
\hat{\cal J} _{LL^{\prime}1} (r) \vert \tilde{i} \rangle, 
\label{stepme}
\end{eqnarray}
\begin{eqnarray*}
j_+ (fi) = V(f,i)  \langle f \vert \hat{J}_+ \vert i \rangle, \\ 
j_+ (f \tilde{i}) = V(f,i) \langle f \vert \hat{J}_+ \vert \tilde{i} 
\rangle .
\end{eqnarray*}
The corresponding matrix elements of $\hat{Q}_{21}$ are defined in
analogy with $\hat{\rho}_{21}$ above, and those of $\hat{L}_+$ and 
$\hat{S}_+$ -- in analogy with $\hat{J}_+$. The symmetries of such
step-up and step-down matrix elements can be derived from
(\ref{qpsym}) using (\ref{mop}). These symmetries allow us to
manipulate the microscopic sums in (\ref{intme}) in order to get rid
of the redundant terms. Upon insertion in (\ref{rmel}), one arrives
to the the following final expression for the r.m.e.  (\ref{rmel}) in
terms of step-up matrix elements:
\begin{eqnarray}
\langle L^{\pi}1, \nu \parallel \hat{T}^{\dag}_L  \parallel 0, 
\text{g.s.}  \rangle 
= {1 \over 2} \sum _{0 < i < f} \Bigl [ F^{\nu} _{\sigma} (fi) 
b_{\sigma} (fi,r) 
\nonumber \\
+  F^{\nu} _{\sigma} (f\tilde{i}) b_{\sigma} (f\tilde{i},r) 
\Bigr ], \label{redsum} 
\end{eqnarray}
where $b_{\sigma} (fi,r)$, $b_{\sigma} (f\tilde{i},r)$ are the
step-up matrix elements (\ref{stepme}) of the multipole transition
density operators. The notation $0 < i < f$ indicates that the final
state $f$ has a larger $K$-value than the initial state $i$.  The
amplitudes of the QRPA phonon wave function are contained in the new
step-up factors $F^{\nu} _{\sigma}$ (\ref{redsum}), which are related
with the old ones (\ref{fsig}) through:
\begin{eqnarray}
 F^{\nu} _{\sigma} (fi) = \sqrt{2} \Bigl [  F^{\nu} _{\sigma} (fi,
 -1) -  F^{\nu} _{\sigma} (fi,  +1) \Bigr ], \nonumber \\
 F^{\nu} _{\sigma} (f\tilde{i}) = \sqrt{2} \Bigl [ F^{\nu} 
 _{\sigma} (f\tilde{i}, +1) -  F^{\nu} _{\sigma} (f\tilde{i}, -1) 
 \Bigr ], \nonumber \\
  = - \sqrt{2} \Bigl [ F^{\nu} _{\sigma} (\tilde{f}i, -1) + F^{\nu} 
  _{\sigma} (\tilde{f}i,  +1)  \Bigr ], \label{fup}
\end{eqnarray}
i.e., they are obtained from the old factors (\ref{fsig}) by simply
replacing q.p. matrix elements of $m$-symmetrized operators with
those (\ref{stepme}) of the corresponding step-up operators. 

\section{Matrix elements of transition densities}

We are going to derive expressions for the matrix elements
(\ref{stepme}) of the transition density operators $\hat{\rho} _{21}
(r)$ and $\hat{\cal J} _{LL^{\prime}1} (r)$ (\ref{mdop}). They are
necessary for the calculation of the $E2$ and $M1$ transition densities
(\ref{rhoden}), (\ref{jden}). The vector scalar product in
(\ref{mdop}) is expressed by the covariant components $T_{\mu}$ of
the involved vectors \cite{varsh}:
\begin{eqnarray*}
[ {\bf Y} _{LL^{\prime}M} (\Omega ) ] _{\mu} = (-1) ^{\mu}
(L^{\prime}, M+\mu , 1, -\mu \vert LM)  \\
\times Y_{L^{\prime}, M+\mu} (\Omega ),  \\
 {\bf Y} _{LL^{\prime}M} (\Omega ) \bullet {\bf T}_1 = \sum_{\mu} 
 (L^{\prime}, M+\mu , 1, -\mu \vert LM) \\ 
 \times  Y_{L^{\prime}, M+\mu} (\Omega ) T_{1,-\mu}, 
\end{eqnarray*}
\begin{eqnarray}
 {\bf Y} _{LL^{\prime}M} (\Omega ) \bullet [ \bbox{\nabla}_1 \times 
 {\bf T}_1 ] _1 \nonumber \\ 
 = -i \sqrt{2} \sum _{\mu\nu\lambda} 
 (L^{\prime}, M+\mu , 1, -\mu \vert LM) (1,\nu , 1, \lambda \vert 
 1\mu) \nonumber \\
 \times Y_{L^{\prime}, M+\mu} (\Omega ) \nabla _{1\nu} T_{1,\lambda},  
\label{scalar} 
\end{eqnarray}
where the round brackets denote the Clebsch-Gordan coefficients. 

The action of $\phi$-dependent operators on the basis functions $\psi
(\Lambda , \phi )$ (\ref{basis}) and spin components on the spinors
$\chi (\Sigma )$ is calculated analytically. Thus, the integration
over the azymuth angle $\phi$ and the spin variables, performed
analytically, results in selection rules for the quantum numbers
$\Lambda , \, \Sigma$ of the involved Woods-Saxon initial and final
wave functions $(f,i)$ (\ref{wswf}).  Therefore, only the first two
basis functions, i.e., the Hermite and Laguerre polynomials from
(\ref{basis}), remain in the final expressions, which have to be
integrated over the polar angle $\theta$.  Due to the different
selection rules, we treat explicitly the two different terms with
$\Lambda$, $\Lambda^{\prime}$ in the Woods-Saxon eigenfunctions
(\ref{wswf}), (\ref{twswf}). Let us introduce a shorter notation
$A^{fi}$ for the product of these two functions, belonging to the
initial and final Woods-Saxon states (\ref{wswf}), involved in the
considered matrix element $(f,i)$. The derivatives of such an
expression with respect to $\rho$ and $z$ can also be written
afterwards in a compact form:
\begin{mathletters} \label{der}
\begin{eqnarray}
A^{fi} (\Lambda_f, \Lambda_i, \rho ,z) = \sum_{kj} C^f_k 
(\Lambda_f) C^i_j (\Lambda_i) \nonumber  \\ 
 \times \psi ^{\Lambda_f}_k (\rho ) \, \psi ^{\Lambda_i}_j (\rho ) 
 \> \psi_k (z) \psi_j (z), \label{afi} \\ 
B^{fi} (\Lambda_f, \Lambda_i, \rho ,z) = {\partial \over \partial
\rho_i } A^{fi} - { \vert \Lambda_i \vert \over \rho } A^{fi}  
\nonumber \\
= \sum_{kj} C^f_k (\Lambda_f) C^i_j (\Lambda_i) \> \psi ^{\Lambda_f}_k 
(\rho ) \, D^{\Lambda_i}_j (\rho ) \> \psi_k (z) \psi_j (z), \nonumber
\end{eqnarray}
\begin{eqnarray}
C^{fi} (\Lambda_f, \Lambda_i, \rho ,z) = {\partial \over \partial
z_i } A^{fi} (\Lambda_f, \Lambda_i, \rho ,z) \nonumber \\
= \sum_{kj} C^f_k (\Lambda_f) C^i_j (\Lambda_i) \> \psi ^{\Lambda_f}_k 
(\rho ) \, \psi ^{\Lambda_i}_j (\rho ) \> \psi_k (z) D_j (z), 
\label{bcder}
\end{eqnarray}
\begin{eqnarray*}
R^{fi} (\Lambda_f, \Lambda_i, \rho ,z) = {\partial \over \partial
\rho} A^{fi}  = \sum_{kj} C^f_k (\Lambda_f) C^i_j (\Lambda_i) \\
\times \Bigl [ \psi ^{\Lambda_f}_k (\rho ) \, D^{\Lambda_i} 
_j (\rho ) + D^{\Lambda_f}_k (\rho ) \, \psi ^{\Lambda_i}_j (\rho ) 
 \Bigr ] \psi_k (z) \psi_j (z) \\ 
 + { (\vert \Lambda_f \vert + \vert \Lambda_i \vert ) \over \rho } 
 A^{fi}  (\Lambda_f, \Lambda_i, \rho ,z), 
\end{eqnarray*}
\begin{eqnarray}
Z^{fi} (\Lambda_f, \Lambda_i, \rho ,z) = {\partial \over \partial z} 
A^{fi} (\Lambda_f, \Lambda_i, \rho ,z) \nonumber \\
= \sum_{kj} C^f_k (\Lambda_f) C^i_j (\Lambda_i) \> 
\psi ^{\Lambda_f}_k (\rho ) \, \psi ^{\Lambda_i}_j (\rho ) 
\nonumber \\
\times \bigl [ \psi_k (z) D_j (z) + D_k (z) \psi_j (z) \bigr ].
\label{rzder}
\end{eqnarray}
\end{mathletters}
\noindent
The factors $C^i_j, \, C^f_k$ in the above expressions are the
expansion coefficients (\ref{wswf}) of the initial and final
Woods-Saxon states, respectively. The symbolic notations 
$\partial /\partial \rho_i$ and $\partial /\partial z_i$, appearing
in (\ref{bcder}), indicate that only the basis function belonging to
the initial state is subjected to differentiation. It is seen from
(\ref{bcder}), (\ref{rzder}) that all derivatives are expressed
through the elementary derivatives $D^{\Lambda}_j (\rho )$ and $D_j
(z)$ (\ref{deriv}) with respect to $\rho$ and $z$. 

Using the above short notations for derivatives and (\ref{scalar})
for the vector scalar product, one can write the q.p. matrix elements
(\ref{stepme}) of the multipole transition densities (\ref{rhoden}),
(\ref{jden}) in a more compact form. The matrix elements of the charge
transition density (\ref{rhoden}) are given by:
\begin{eqnarray*}
\rho _{21} (fi,r) = e \varepsilon U(f,i) \int Y_{21} (\Omega ) \Phi
^{\dag} _f ({\bf r}) \Phi _i ({\bf r}) d \Omega  \\ 
= -e \varepsilon U(f,i) \sqrt{ 15 \over 8 \pi} \int {\rho z \over r^2} 
\\
\times \bigr [ A^{fi} (\Lambda +1, \Lambda , \rho ,z) + 
 A^{fi} (\Lambda +2, \Lambda +1, \rho ,z) \bigr ] d \> {\rm cos} 
 \theta , \\
\rho = r \> {\rm sin} \theta , \quad z = r \> {\rm cos} \theta , 
\end{eqnarray*}
\begin{eqnarray}
\rho _{21} (f\tilde{i},r) = -e \varepsilon U(f,i) \sqrt{ 15 \over 8 
\pi} \int {\rho z \over r^2} \nonumber \\
\times \bigr [ A^{fi} (1, 0, \rho ,z) -  A^{fi} (0, -1, \rho ,z) 
\bigr ] d \> {\rm cos} \theta , \label{rhome}
\end{eqnarray}
where $\Phi ({\bf r})$ are the Woods-Saxon wave functions
(\ref{wswf}). The integration over the polar angle in (\ref{rhome})
and similar expressions below is limited to the interval $0 \le
\theta \le \pi /2$ (or $0 \le {\rm cos} \, \theta \le 1$) due to
reflection symmetry of the basis with respect to the plane $z=0$.

The matrix elements (\ref{stepme}) of the transverse (current)
multipole density operators $\hat{\cal J} _{LL^{\prime}1} (r)$
(\ref{mdop}) can be decomposed into convection and spin parts,
originating from the convection and magnetization current density
operators ${\bf j}^C ({\bf r})$ (\ref{convop}) and ${\bf j}^S ({\bf
r})$ (\ref{spinop}),
\begin{equation}
{\cal J} _{LL^{\prime}1} (fi,r) =  {\cal J}^C _{LL^{\prime}1} (fi,r)
+  {\cal J}^S _{LL^{\prime}1} (fi,r), \label{jme} 
\end{equation}
and the same for ${\cal J} _{LL^{\prime}1} (f\tilde{i},r)$. All of
them are purely imaginary in our basis, see Table \ref{tabphas}.
They are necessary for the calculation of the the transverse
(current) transition densities (\ref{jden}).  The matrix elements
(\ref{jme}), obtained from (\ref{scalar}) and expressed in nuclear
magnetons $\mu_N = e \hbar /(2Mc) \approx 0.1052 \; e\,$fm, have
the form: 
  \vfill \eject
     \widetext \onecolumn 
\begin{eqnarray}
{\cal J}^C _{LL^{\prime}1} (fi,r) = -2i \varepsilon \mu _N V(f,i) 
\sum_{\mu} (L^{\prime}, 1+\mu , 1, -\mu \vert L1)  
\int  Y_{L^{\prime}, 1+\mu} (\Omega )  \Phi^{\dag} _f ({\bf r}) 
 \bigl [ \hat{\nabla} _{1,-\mu} \Phi _i ({\bf r}) \bigr ] d \Omega , 
\label{jc} \\ 
{\cal J}^S _{LL^{\prime}1} (fi,r) = - i \mu _N g^S V(f,i) \sqrt{2} 
\sum _{\mu\nu\lambda} (L^{\prime}, 1+\mu , 1, -\mu \vert L1) 
(1,\nu , 1, \lambda \vert  1, -\mu) \int  
Y_{L^{\prime}, 1+\mu} (\Omega ) \hat{\nabla} _{1\nu} \bigl [  
 \Phi^{\dag} _f ({\bf r}) \hat{S} _{1,\lambda} \Phi _i ({\bf r}) 
 \bigr ] d \Omega . \label{js}
\end{eqnarray}

   One obtains from (\ref{jc}) the convection matrix elements of the
transverse $M1$ current with $LL^{\prime} = $11 and $E2$ currents with
$LL^{\prime} = $21, 23:
\begin{mathletters} \label{jcl}
\begin{eqnarray}
{\cal J}^C _{111} (fi,r) = - i \varepsilon \mu _N V(f,i) \sqrt{
3 \over 4 \pi}  
\sum _{\Lambda^{\prime} = \Lambda, \Lambda +1} \int \biggl \{ 
{z \over r} \Bigl [ B^{fi} (\Lambda^{\prime} +1, \Lambda^{\prime}, 
\rho , z) 
+ {( \vert \Lambda^{\prime} \vert - \Lambda^{\prime}) \over \rho }
 A^{fi} (\Lambda^{\prime} +1, \Lambda^{\prime}, \rho , z) \Bigr ] 
 \nonumber \\ 
- {\rho \over r} C^{fi} (\Lambda^{\prime} +1, \Lambda ^{\prime}, 
\rho , z) \biggr \}  d \> {\rm cos} \theta , 
\label{jc11}
\end{eqnarray}
\begin{eqnarray}
{\cal J}^C _{211} (fi,r) = i \varepsilon \mu _N V(f,i) \sqrt{
3 \over 4 \pi}  
\sum _{\Lambda^{\prime} = \Lambda, \Lambda +1} \int \biggl \{ 
{z \over r} \Bigl [ B^{fi} (\Lambda^{\prime} +1, \Lambda^{\prime}, 
\rho , z) 
+ {( \vert \Lambda^{\prime} \vert - \Lambda^{\prime}) \over \rho }
 A^{fi} (\Lambda^{\prime} +1, \Lambda^{\prime}, \rho , z) \Bigr ] 
 \nonumber \\ 
+ {\rho \over r} C^{fi} (\Lambda^{\prime} +1, \Lambda ^{\prime}, 
\rho , z) \biggr \}  d \> {\rm cos} \theta , 
\label{jc21}
\end{eqnarray}
\begin{eqnarray}
{\cal J}^C _{231} (fi,r) = {i \varepsilon \mu _N \over \sqrt{2\pi}} 
V(f,i) \sum _{\Lambda^{\prime} = \Lambda, \Lambda +1} 
\int \biggl \{ { (z^2 - 4 \rho ^2) \over r^3 } \Bigl [   
 z  B^{fi} (\Lambda^{\prime} +1, \Lambda ^{\prime}, \rho , z)  
+ \rho C^{fi} (\Lambda^{\prime} +1, \Lambda ^{\prime}, \rho , z) 
\Bigr ]  \nonumber \\ 
- { z \Lambda^{\prime} \over \rho r} A^{fi} (\Lambda ^{\prime} +1, 
\Lambda ^{\prime}, \rho , z) \biggr \} d \> {\rm cos} \theta ,  
\label{jc23}
\end{eqnarray} 
\end{mathletters}
\noindent
where $\Lambda = K_i - 1/2$ in (\ref{jcl}) is determined by the
$K$-number of the initial state $K_i$. Let us note that here and
below we have always $K_f = K_i +1$.  Although $\Lambda \ge 0$ in the
above two-term sums over $\Lambda ^{\prime}$, the expressions  are
written in a more general way in order to use them also for the
corresponding matrix elements ${\cal J}^C _{LL^{\prime}1}
(f\tilde{i},r)$,  involving a time-reversed initial state. They are
obtained from (\ref{jcl}) by putting $\Lambda = K_i - 1/2 = -1$ and
multiplying the first term ($\Lambda ^{\prime} = -1$) with an overall
factor --1, i.e., the two-term sums run over $\Lambda ^{\prime} = -1,
0$ in this special case. The same structure was written explicitly in
(\ref{rhome}) for $\rho _{21} (f\tilde{i},r)$.  The phase --1
originates from the second term of the time-reversed Wood-Saxon wave
function (\ref{twswf}).

     The spin matrix elements of the transverse $M1$ and $E2$ currents, 
obtained from (\ref{js}), have the form:
\begin{mathletters} \label{jsl}
\begin{eqnarray}
{\cal J}^S _{111} (fi,r) = -i g^S \mu _N V(f,i) {1 \over 8} \sqrt{3 
\over \pi}  \int \biggl \{ {z \over r} \Bigl [ R^{fi} (\Lambda  +1, 
\Lambda , \rho , z) -  R^{fi} (\Lambda  +2, \Lambda +1, \rho , z) 
\nonumber \\
+ { 1 \over \rho } \bigl [ A^{fi} (\Lambda  +1, \Lambda , \rho , z) 
- A^{fi} (\Lambda  +2, \Lambda +1, \rho ,z) \bigr ] 
- 2  Z^{fi} (\Lambda  +1, \Lambda +1, \rho , z) \Bigr ] 
\nonumber \\
+ {\rho \over r} \Bigl [  R^{fi} (\Lambda  +2, \Lambda , \rho , z) 
- R^{fi} (\Lambda  +1, \Lambda +1, \rho , z) + { 2 \over \rho } 
A^{fi} (\Lambda  +2, \Lambda , \rho ,z) \Bigr ] \biggr \} d \> 
{\rm cos} \theta , \label{js11}
\end{eqnarray}
\begin{eqnarray}
{\cal J}^S _{211} (fi,r) = i g^S \mu _N V(f,i) {1 \over 8} \sqrt{3 
\over \pi}  \int \biggl \{ {z \over r} \Bigl [ R^{fi} (\Lambda  +1, 
\Lambda , \rho , z) -  R^{fi} (\Lambda  +2, \Lambda +1, \rho , z) 
\nonumber \\
+ { 1 \over \rho } \bigl [ A^{fi} (\Lambda  +2, \Lambda +1, \rho ,z)
-  A^{fi} (\Lambda  +1, \Lambda , \rho , z) \bigr ] 
+ 2  Z^{fi} (\Lambda  +1, \Lambda +1, \rho , z) \Bigr ] 
\nonumber \\
+ {\rho \over r} \Bigl [  R^{fi} (\Lambda  +1, \Lambda +1, \rho , z) 
- R^{fi} (\Lambda  +2, \Lambda , \rho , z) + { 2 \over \rho } 
A^{fi} (\Lambda  +2, \Lambda , \rho ,z) \Bigr ] \biggr \} d \> 
{\rm cos} \theta , \label{js21}
\end{eqnarray}
   \vfill \eject \vbox{\vfill \eject}
\begin{eqnarray}
{\cal J}^S _{231} (fi,r) = {i g^S \mu _N \over 4 \sqrt{2 \pi} } 
V(f,i) \int \biggl \{  {z \over r}  \Bigl [ R^{fi} (\Lambda +1, 
\Lambda , \rho , z) - R^{fi} (\Lambda +2, \Lambda +1, \rho , z) 
\Bigr ] \nonumber \\
+ { (z^2 - 4 \rho ^2 ) \over r^3 }  \Bigl [ \rho \bigl [ 
R^{fi} (\Lambda +2, \Lambda , \rho , z) - R^{fi} (\Lambda +1, 
\Lambda +1, \rho , z) \bigr ] - 2 A^{fi} (\Lambda  +2, \Lambda , 
\rho ,z)  \nonumber \\
+  { z \over \rho }  \bigl [ A^{fi} (\Lambda  +1, 
\Lambda , \rho ,z) - A^{fi} (\Lambda  +2, \Lambda +1, \rho ,z) 
\bigr ] \Bigl ] \nonumber \\
- { z \over r^3 } \Bigl [ \bigl ( 2z^2 - 3 \rho ^2 \bigr )  
Z^{fi} (\Lambda  +1, \Lambda +1, \rho , z)  - 5 \rho ^2 
Z^{fi} (\Lambda +2, \Lambda , \rho , z) \Bigr ]  \biggr \} 
d \> {\rm cos} \theta , \label{js23}
\end{eqnarray}
\end{mathletters}   
                      \narrowtext
The two terms with $\Lambda^{\prime} = \Lambda , \Lambda +1$ from
(\ref{jcl}) are written explicitly in the expressions (\ref{jsl})
because their symmetry is lost in the spin matrix elements.
Nevertheless, one can still use (\ref{jsl}) also for the
corresponding matrix elements ${\cal J}^S _{LL^{\prime}1}
(f\tilde{i},r)$. They are obtained from (\ref{jsl}) by putting
$\Lambda = K_i - 1/2 = -1$ and multiplying with --1 each functional
$A, \, R,$ or $Z$  (\ref{der}), which has --1 as a second argument.
The radius $r$ in the r.h.s. of (\ref{jcl}), (\ref{jsl}) is a
constant with respect to the integration. Thus, $\rho$ and $z$ are
not independent integration variables but simple functions of the
integration angle $\theta$, given in (\ref{rhome}).

   \vfill \eject \twocolumn
%
%

%
%
\narrowtext
%
%
\begin{table} 
\caption{Comparison with experiment of QRPA excitation energies
$E_x$, upwards ($\uparrow$) $M1$ and $E2$ transition probabilities
and orbit/spin ratios $R_{\rm o.s.}$  \protect\cite{nofa88} of the
$M1$ matrix elements for the strongest orbital $K^{\pi} =1^+$
excitation in each rare-earth nucleus with experimentally measured
$(e,e^{\prime})$ cross section of this excitation
\protect\cite{bohle84,hart89,boh84}.}
\label{tabrpa}

\begin{tabular}{llllll}  
  
Nucleus & exp.$^a$ or & $E_x$ & $B(M1)$ & $B(E2) \,^b$ & 
$R_{\rm o.s.} \, ^c$ \\  
        & theor.  & (MeV)  & ($\mu^2_N$) & (e$^2$fm$^4$) &  \\
\tableline  
\vspace{2pt}
$^{154}$Sm & exp.$^d$ & 3.20 & 0.80(20) &       & $> 0.8$ \\
           & th.      & 3.16 & 0.87     & 28    & 7.3     \\
$^{154}$Gd & exp.$^e$ & 2.94 & 0.85(9)  &       &         \\
           & th.      & 2.94 & 1.00     & 30    & 6.8     \\
$^{156}$Gd & exp.$^f$ & 3.07 & 1.30(20) & 40(6) & $> 1.3$ \\
           & th.      & 2.90 & 1.24     & 42    & 7.8     \\
$^{158}$Gd & exp.$^g$ & 3.20 & 0.77(9)  &       &         \\
           & th.      & 3.48 & 0.89     & 37    & 7.0     \\
$^{168}$Er & exp.$^d$ & 3.39 & 0.90(20) &       &         \\
           & th.      & 3.35 & 0.88     & 18    & 9.6     \\
\end{tabular} 

\noindent $^a$ A reference for the experimental $E_x$ and $B(M1)$
values is given in this column. \\
\noindent $^b$ Experimental data are available only for $^{156}$Gd 
\cite{bohle85}. \\ 
\noindent $^c$ Experimental data from $(e,e^{\prime})$ and 
$(p,p^{\prime})$ \cite{dja85}. \\ 
\noindent $^d$ Ref. \cite{boh84}. \\
\noindent $^e$ Ref. \cite{hart89}. \\
\noindent $^f$ Ref. \cite{bohle86}. \\
\noindent $^g$ Ref. \cite{pitz89}. \\

\end{table}   

%
\begin{table} 

\caption{Phases c$^{\rm H}$, c$^{\rm T}$ (\protect\ref{phase}) of the
$m$-symmetrized operators (\protect\ref{mop}) under hermitian and
time conjugation, respectively. c$^{\rm C}$ is the phase of the
matrix elements in the basis
(\protect\ref{basis}),(\protect\ref{wswf}),(\protect\ref{twswf}). }
\label{tabphas}
\begin{tabular}{lccc}  
  
Phase & $\hat{Q}(m), \; \hat{\rho} _2 (m,r)$  & $\hat{\cal J} 
_{LL^{\prime}} (m,r)$ & $\hat{J}(m), \> \hat{L}(m), \> \hat{S}(m)$ \\ 
\tableline  

c$^{\rm H}$ & $+m$ & $+m$ & $-m$ \\  
c$^{\rm T}$ & $+m$ & $-m$ & $+m$ \\
c$^{\rm C}$ & $+1$ & $-1$ & $+1$ \\
\end{tabular} 
\end{table}
%
%
%
\begin{figure} 
\caption{QRPA transition densities (\protect\ref{den}) for $M1$
($L=1$) and $E2$ ($L=2$) excitations of the $K^{\pi} = 1^+$ state
from Table\ \protect\ref{tabrpa} in $^{156}$Gd.  The total
(continuous curve) transverse transition density ${\cal J}^{\nu}
_{LL^{\prime}} (r)$  is a sum of the proton convection (long dashed
curve) and magnetization (short dashed curve for neutrons and dotted
curve for protons) transition densities.} \label{fig1}
\end{figure}

%
\begin{figure}
\caption{The same as in Fig.\ \protect\ref{fig1} but for $^{168}$Er.}
\label{fig2}
\end{figure}

%
\begin{figure}
\caption{DWBA $(e,e^{\prime})$ cross sections for $M1$ (dashed
curves) and $E2$ (dotted curves) excitations, corresponding to the
QRPA state with $K^{\pi} =1^+$ from Table\ \protect\ref{tabrpa} in
$^{154}$Sm (top plot) and  $^{154}$Gd (bottom plot). Scattering angle
$\theta = 165 ^{\circ}$. For each nucleus the sum of the $M1$ and
$E2$ cross sections (continuous curve), plotted versus the effective
momentum transfer $q_{\rm eff}$ \protect\cite{fanota}, is compared to
experimental data (dots with error bars) from \protect\cite{boh84}
and \protect\cite{hart89}, respectively.} \label{fig3}
\end{figure}

%
\begin{figure}
\caption{The same as in Fig.\ \protect\ref{fig3} but for
$^{156,158}$Gd with experimental data from
\protect\cite{bohle84,boh84}, respectively.  The DWBA cross sections
for $^{156}$Gd are obtained with the QRPA transition densities from
Fig.\ \protect\ref{fig1}.}  \label{fig4}
\end{figure}

%
\begin{figure}
\caption{The same as in Fig.\ \protect\ref{fig3} but for $^{168}$Er
with experimental data from \protect\cite{boh84}. The DWBA cross
sections are obtained with the QRPA transition densities from Fig.\
\protect\ref{fig2}.}
\label{fig5}
\end{figure}

%
\begin{figure}
\caption{The theoretical $M1$ cross section for $^{156}$Gd from the
upper plot of Fig.\ \protect\ref{fig4} is plotted here as a
continuous curve up to a higher momentum transfer $q_{\rm eff}$. 
Dashed curve: the contribution from the proton
convection transition density ${\cal J} ^{\nu ,C} _{11} (r)$ 
(\protect\ref{jden}), (\protect\ref{jc11}) alone.  These DWBA results
are compared also to the total (convection plus spin) $M1$ cross
section in PWBA (dot-dashed curve).}
\label{fig6}
\end{figure}

%
\begin{figure}
\caption{Decomposition of the theoretical PWBA electric cross section
(continuous curve: total $E2$) into a sum of its longitudinal $C2$
(dashed curve) and transverse $E2$ (dotted curve) components.
They correspond to the $E2$ excitation in $^{156}$Gd, whose DWBA
cross section is shown in the upper plot of Fig.\ \protect\ref{fig4}.
The PWBA cross section is plotted here for scattering angles $\theta
= 165 ^{\circ}, \, 175 ^{\circ}, \, 178.5 ^{\circ}$, and 179.5$
^{\circ}$ up to a higher momentum transfer $q_{\rm eff}$. }
\label{fig7}
\end{figure}

%
\begin{figure}
\caption{The same as in the upper plot of Fig.\ \protect\ref{fig4}, 
$^{156}$Gd, but for a scattering angle $\theta = 178.5 ^{\circ}$.}
\label{fig8}
\end{figure}

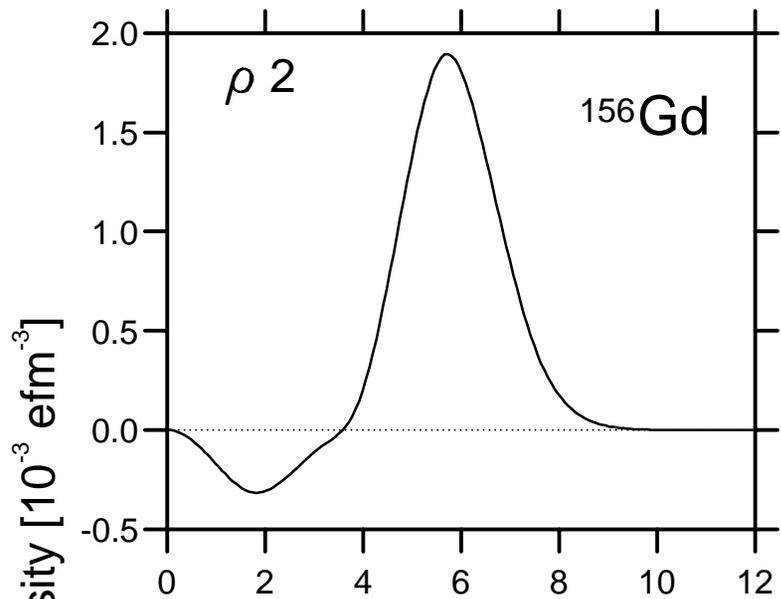
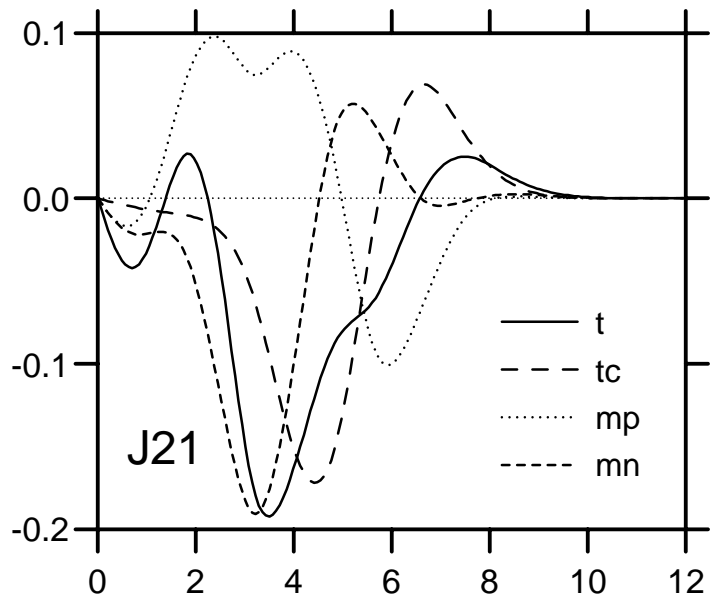
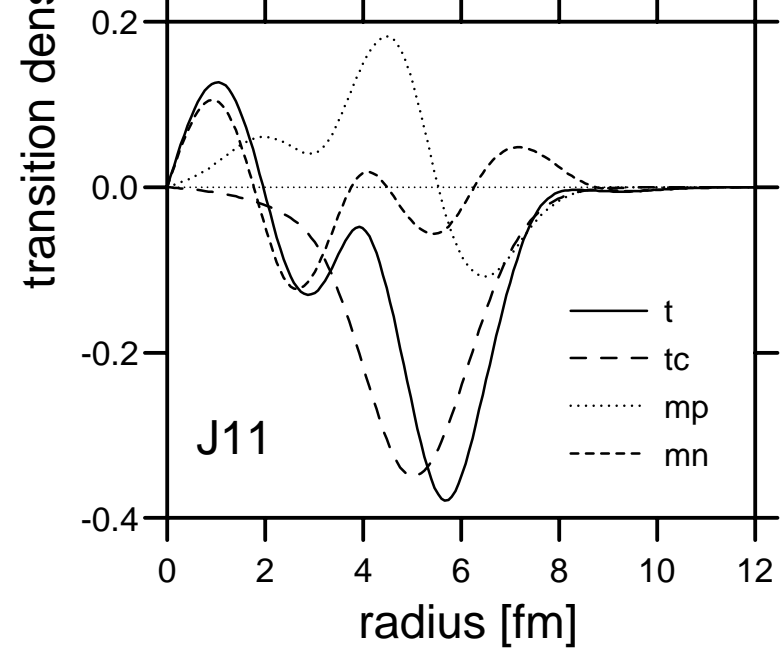
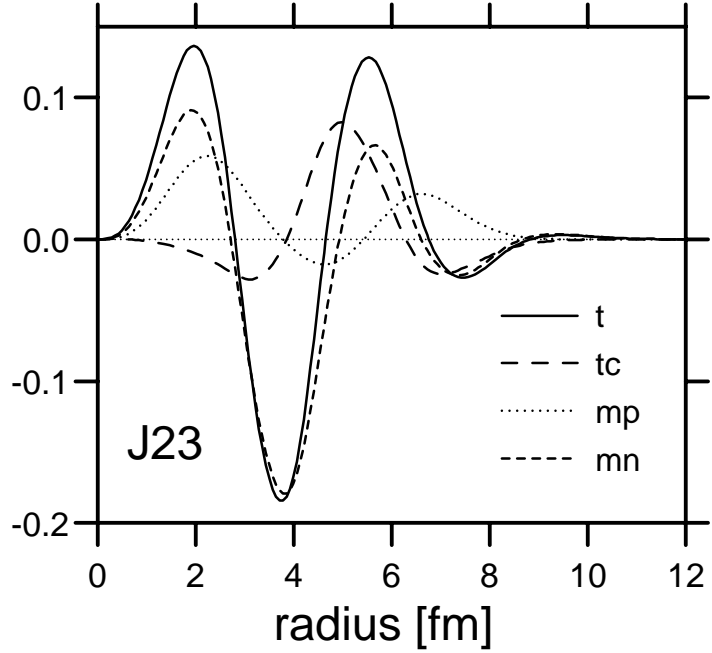
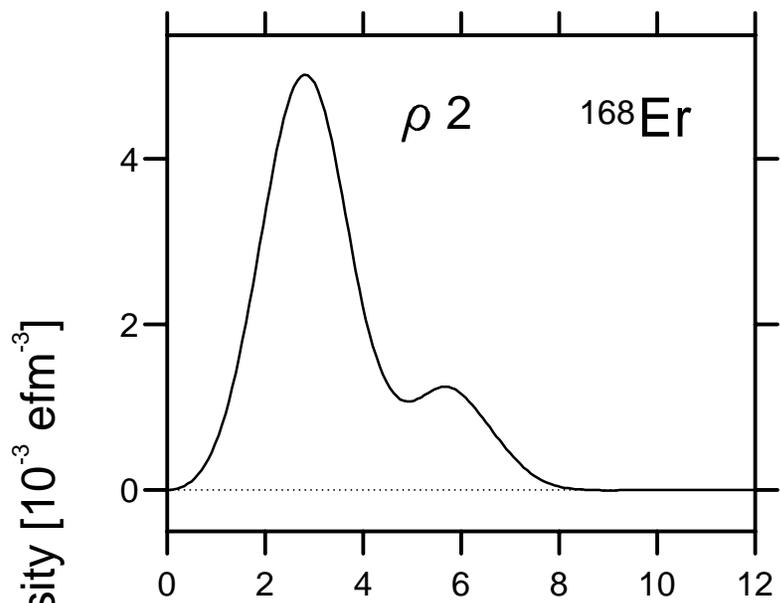
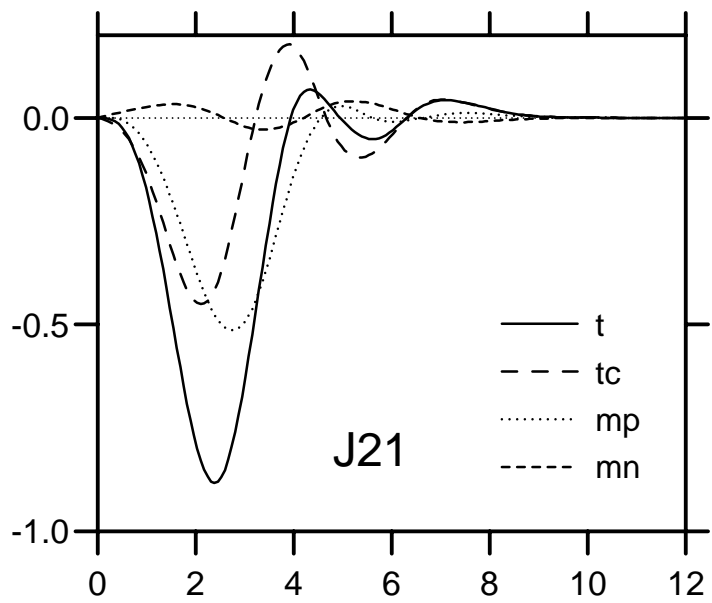
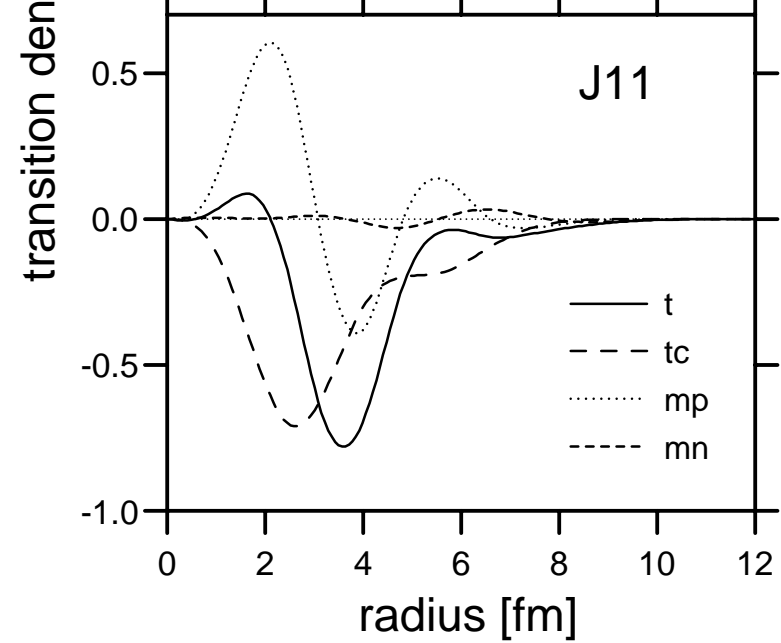
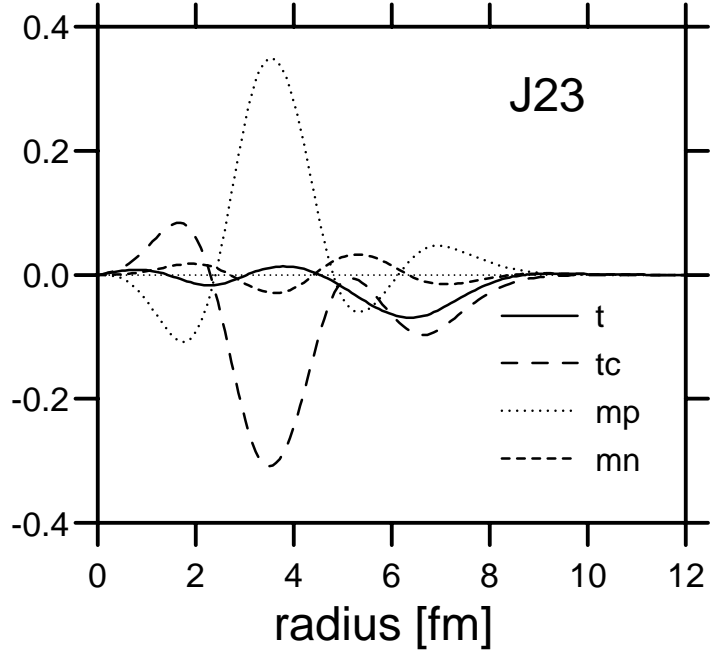
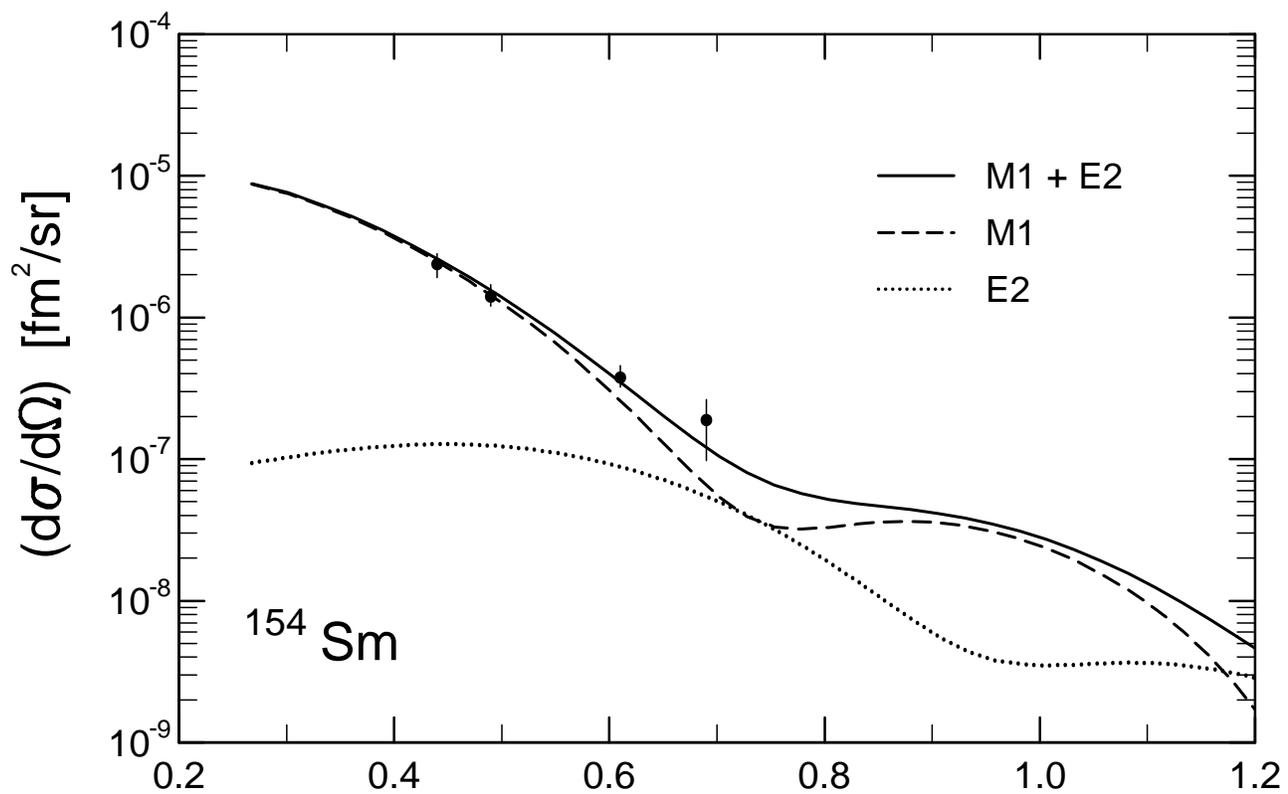
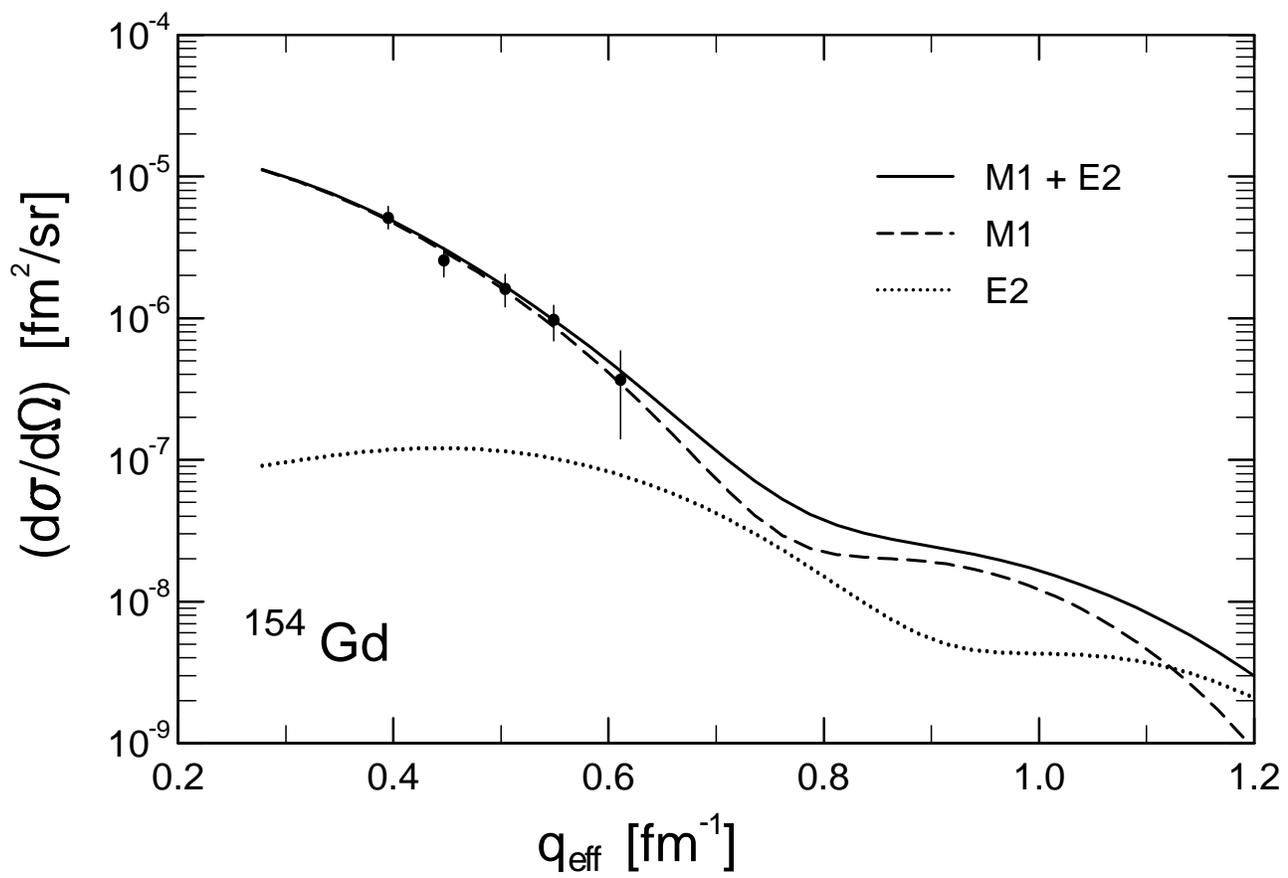
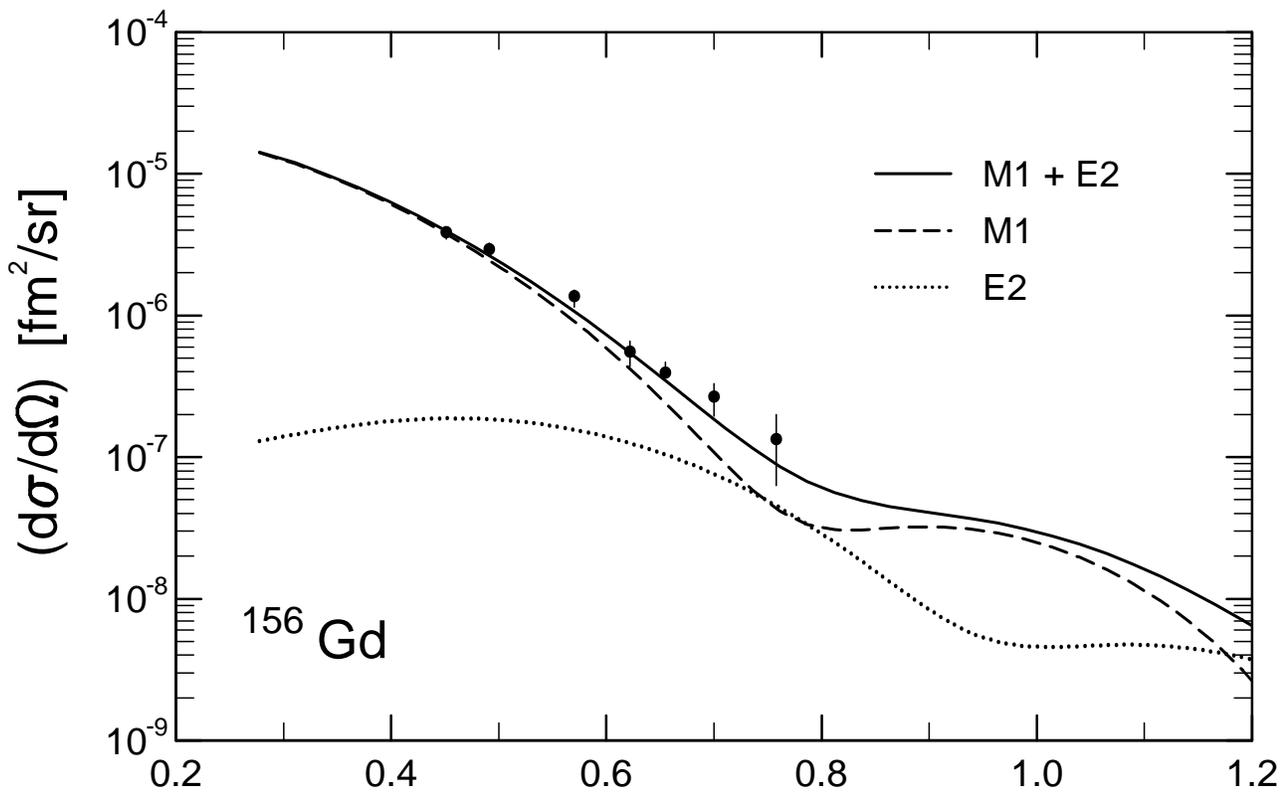
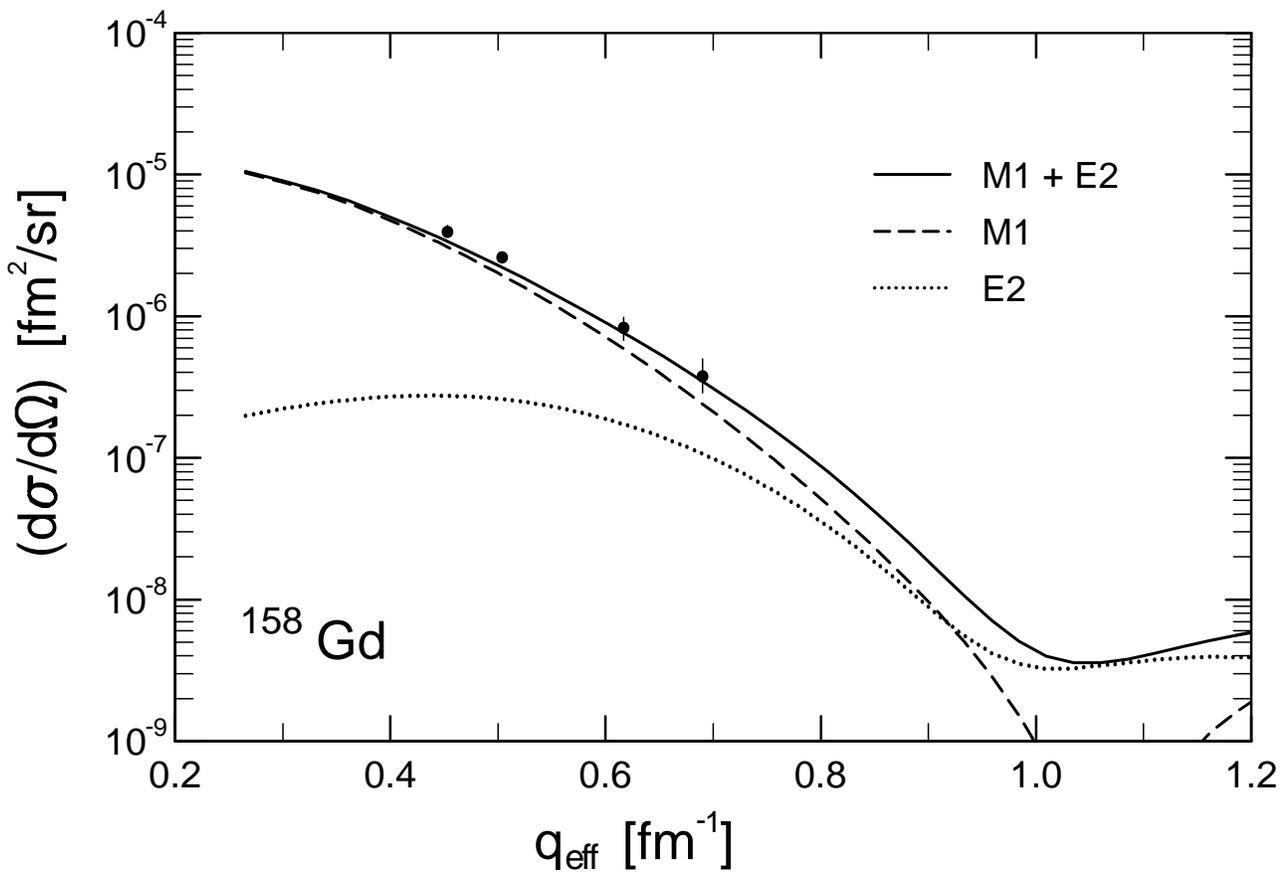
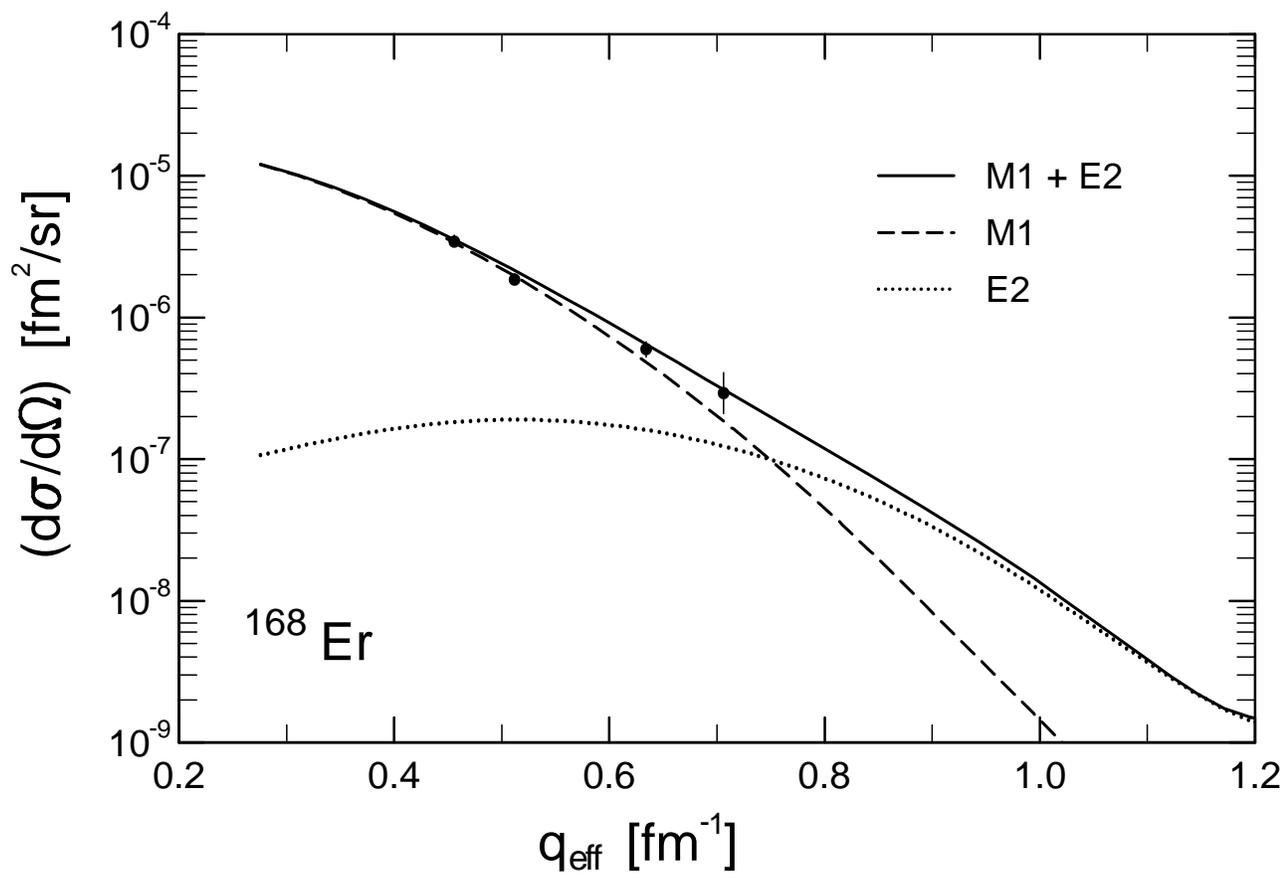
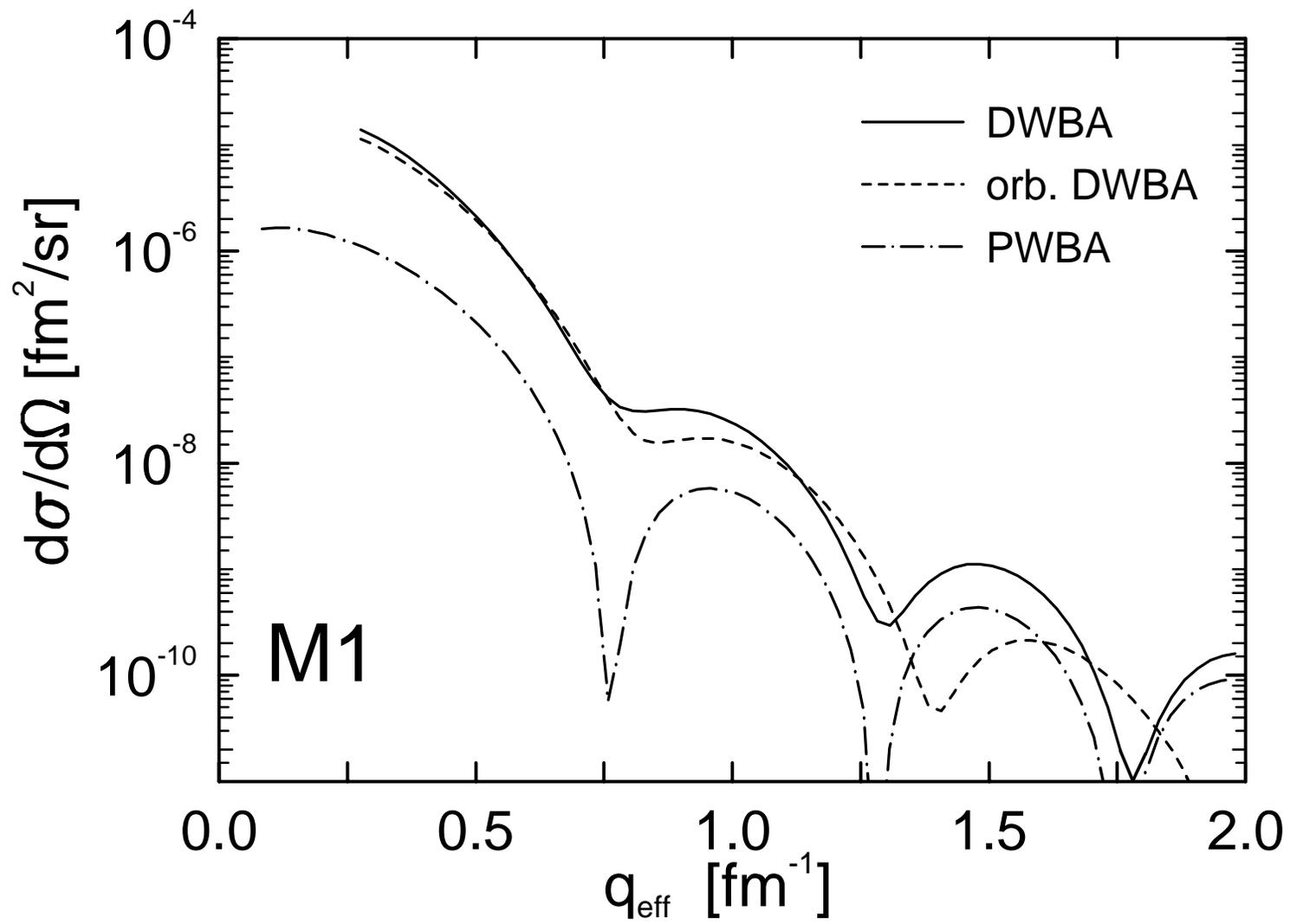
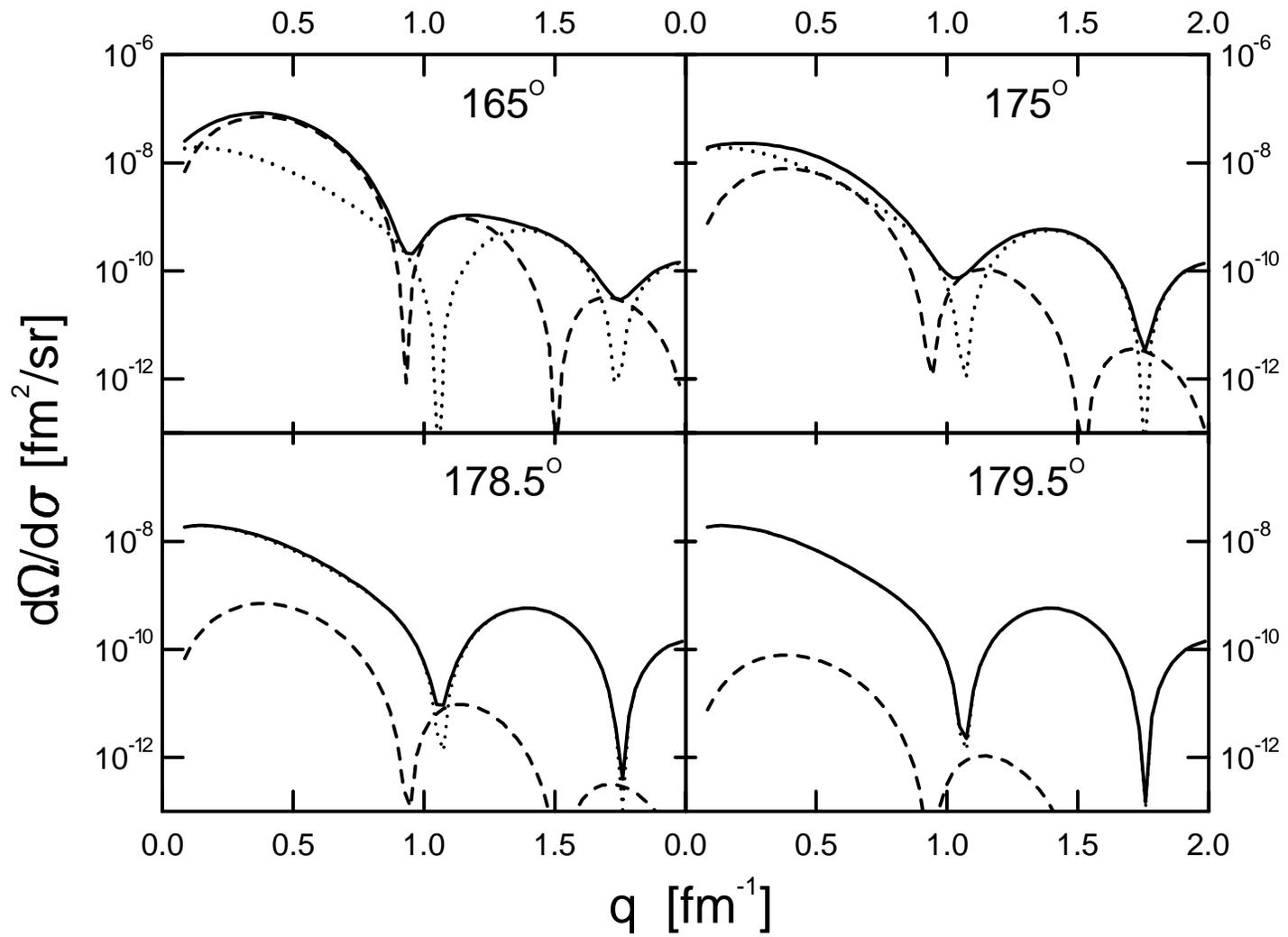
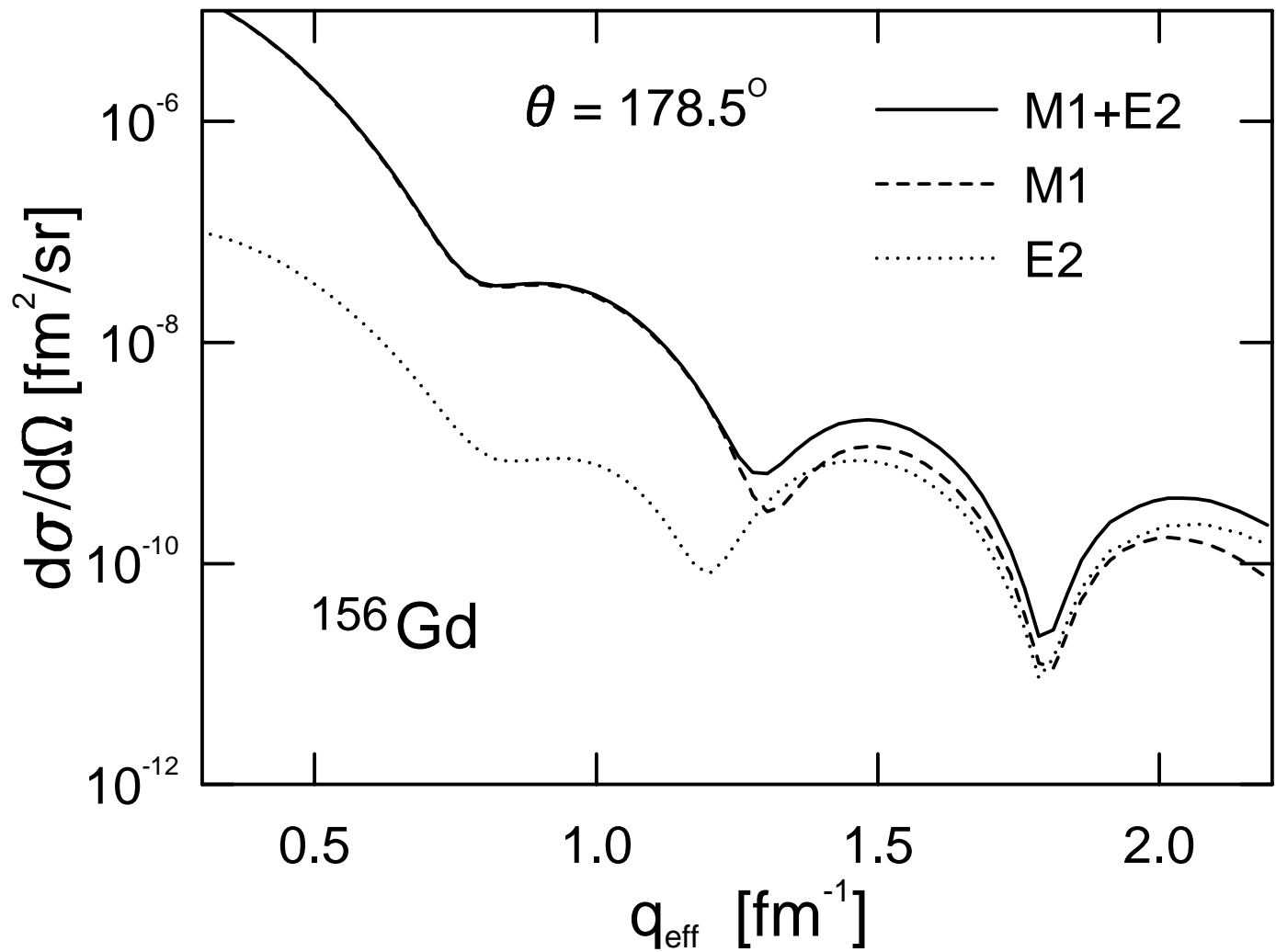
\end{document}